\documentclass[final,5p,times,twocolumn]{elsarticle} 

\usepackage{graphicx}
\usepackage{amssymb}
\usepackage{caption}
\usepackage{subcaption}
\usepackage{rotating}
\usepackage{float}
\usepackage{acronym}
\usepackage{booktabs}
\usepackage{titlesec}
\usepackage{amsmath}
\usepackage{tabularx}
\usepackage{longtable}
\usepackage{siunitx} 
\usepackage{xurl}

\usepackage{threeparttable}

\titleformat{\paragraph}
{\normalfont\normalsize\itshape}{\theparagraph}{1em}{}
\titlespacing*{\paragraph}
{0pt}{3.25ex plus 1ex minus .2ex}{1.5ex plus .2ex}

\usepackage{lineno}
\usepackage[super]{nth}
\usepackage{longtable}
\usepackage{tocloft}

\acrodef{AS}{Active Substances}
\acrodef{HRI}{harmonised Risk Indicators}
\acrodef{IACS}{Integrated Administration and Control System}
\acrodef{JRC}{Joint Research Centre}
\acrodef{PPDB}{Pesticide Properties DataBase}
\acrodef{PPP}{Plant Protection Product}
\acrodef{PRI}{Pesticide Risk Indicator}
\acrodef{RPG}{\textit{Registre Parcellaire Graphique}}
\acrodef{SUR}{Sustainable Use of Plant Protection Products Regulation}
\acrodef{SAIO}{Statistics on Agricultural Input and Output}
\acrodef{TFI}{Treatment Frequency Index}

\usepackage{array}
\newcolumntype{L}{>{\centering\arraybackslash}m{3cm}}

\usepackage[colorlinks=true]{hyperref}
\usepackage{cleveref}

\biboptions{authoryear}

\journal{Journal of Environmental Management}

\begin{document}

\begin{frontmatter}

\title{From parcels to people: development of a spatially explicit risk indicator to monitor residential pesticide exposure in agricultural areas}



\author{Francesco Galimberti\texorpdfstring{$^{1}$},, 
Stephanie Bopp\texorpdfstring{$^{1}$},, Alessandro Carletti\texorpdfstring{$^{1}$},, Rui Catarino\texorpdfstring{$^{1}$},, Martin Claverie\texorpdfstring{$^{1}$},, Pietro Florio\texorpdfstring{$^{1}$},, Alessio Ippolito  \texorpdfstring{$^{2}$},, Arwyn Jones\texorpdfstring{$^{1}$},, Flavio Marchetto\texorpdfstring{$^{3}$},, Michael Olvedy\texorpdfstring{$^{1}$},, Alberto Pistocchi\texorpdfstring{$^{1}$},, Astrid Verhegghen\texorpdfstring{$^{1}$},, Marijn Van Der Velde\texorpdfstring{$^{1}$},, Diana Vieira\texorpdfstring{$^{1}$},, 
Rapha\"{e}l d'Andrimont \texorpdfstring{$^{4}$}}
\address{ $^{1}$\quad European Commission, Joint Research Centre (JRC), Ispra , Italy \\
 $^{2}$\quad European Food Safety Authority (EFSA), Parma, Italy \\
 $^{3}$\quad European Chemicals Agency (ECHA), Helsinki, Finland \\
 $^{4}$\quad European Commission, Joint Research Centre (JRC), Brussels , Belgium
 
}

\begin{abstract}
The increase in global pesticide use has mirrored the rising demand for food over the last decades, resulting in a boost in crop yields. However, concerns about the impact of pesticides on biodiversity, ecosystems, and human health, especially for populations residing close to cultivated areas, are growing. This study investigates how exposure and possible risks to residents can be estimated at high spatial granularity based on plant protection product data. The complexities of such analysis were explored in France, where relevant data with sufficient granularity are publicly available. 
Integrating a range of spatial datasets and exposure assessment methodologies, we have developed an indicator to monitor the levels of pesticide risk faced by residents. By spatialising pesticide sales data according to their authorisation on specific crops, we developed a detailed map depicting potential pesticide loads at parcel level across France. This spatial distribution served as the basis for an exposure and risk assessment, modelled following the European Food Safety Authority’s guidelines.
Combining the risk map with population distribution data, we have developed an indicator that allows to monitor patterns in non-dietary exposure to pesticides. Our results show that in France, on average, 13\% of people might be exposed to pesticides due to living in the proximity to treated crops: the exposure is in the lower range for 34\%, moderate range for 40\% and higher range for 25\% of the exposed population. The risk evaluation is based on worst case assumptions and values should not be taken as a regulatory risk assessment but as indicator to use, for example, for monitoring time trends. The purpose of this indicator is to demonstrate the usefulness of a more granular pesticide data to monitor risk reduction strategies. harmonised and high-resolution data can help in identifying regions where to focus more efforts towards sustainable farming.
\end{abstract}

\begin{keyword}
Risk Assessment \sep Spatialisation \sep Plant Protection Products \sep European Union 


\end{keyword}

\end{frontmatter}



\renewcommand{\thetable}{Table \arabic{table}}
\renewcommand{\thefigure}{Fig. \arabic{figure}}
\renewcommand{\figurename}{}
\renewcommand{\tablename}{}

\section{Introduction}
Pesticides are substances intentionally released into the environment to eliminate pests that have a negative impact on crop production such as weeds, insects, fungi, and others. These substances have played a pivotal role in achieving a four-fold increase in agricultural yields over the past 60 years to meet the growing food demand of the global population \cite{Popp2013}. However, their extensive use can lead to adverse effects on biodiversity \citep{mclaughlin1995impact,mesnage2021improving,khurana2023impact}, with far-reaching consequences for terrestrial, aquatic and marine ecosystems \citep{pistocchi2023screening}, as well as negative impacts on human health \citep{EEA,KIM2017525,DEREUMEAUX2020105210}.
Of particular concern is the health of residents living in close proximity to these agricultural areas, since combined exposure to pesticides remains largely unknown \citep{stehle2015pesticide}.
In response to these, the European Commission has set targets to reduce pesticide use and associated risks, including the Farm to Fork and Biodiversity Strategies \citep{European-Commission2023,European-Commission2023b} and Zero Pollution Ambition \citep{candel2023science}. These targets are set against pesticide sales that have been stable in the EU at around 350 thousand tonnes annually from 2011 to 2020 \citep{EUROSTAT2023,EUROSTAT2023b,EUROSTAT2023c}. To improve the sustainable use of plant protection products (PPP), evidence exists that integrated pest management and eliminating PPPs in sensitive areas, e.g., by creating pesticide-free areas near urban locations, can be effective in reducing exposure to residents \citep{martin2023modelling}. 

Currently, harmonised Risk Indicators (HRI, \ref{sec:hri}) are used to monitor risk from pesticide use in the EU. These indicators rely on data from national sales and reporting on emergency authorisations of PPPs \citep{European-Commission2023c,vekemans2023european}. HRI1, in particular, quantifies the amount of pesticide active substances (AS) placed on the market, weighted according to four groups that intend to represent increasing levels of hazard \citep{Global2000}. Nevertheless, there is no indicator able to assess appropriately the use of pesticides at EU level and the associated risks to humans and ecosystems \citep{vekemans2023european}. Particularly, an indicator that would allow monitoring pesticide exposure of residents living near agricultural fields is missing. Similarly, there is no detailed information on crops grown near residential areas nor the percentage of the population affected. 
Monitoring risk to residents living near treated crops would require a comprehensive dataset containing at minimum the records of pesticide application on fields, including the place and time of the application, information on the agri-environmental landscape, and the distribution of population residing in the proximity of agricultural fields. The development of such a dataset is hampered by the lack of available data on pesticide use on crops \citep{Global2000,PAN-Europe}.
While data collection by public authorities does occur in some cases (e.g. for statistical purposes or to control subsidy applications by farmers), the data is often published with low spatial resolution and not harmonised across Member States (MS). At the EU level, Eurostat serves as the main source of harmonised and reliable information through the collection and publication of pesticide sales data \citep{EUROSTAT2023}. However, data on pesticides are provided with granularity of broad chemical groups at MS level. Similarly, reporting of statistical data on pesticide use on crops is not properly harmonised among MSs, which renders this data not suitable for risk monitoring. Efforts in developing a fine scale spatialisation of pesticide use in various EU countries using official surveys, sales data or remote sensing have been made before \citep{jorgensen2019links, martin2023modelling, strassemeyer54, ward2000identifying, galimberti2020estimating}. 
Even though, in the future, this data scarcity will be partially addressed by the Statistics on Agricultural Input and Output Regulation (SAIO) \citep{EU-Parliament} these changes will still likely provide aggregated pesticide use data that will not allow precise in-situ risk calculations, including proper assessments of risks from pesticide mixtures. From 2026, professional users will have to record PPP use electronically allowing national authorities to exploit this data for better monitoring and risk reduction strategies.

To address these limitations, we propose to create an indicator monitoring the pesticide risk for residents living near crops by modelling the spatial application of pesticides across France, using parcel-scale land use data and information on PPP sales and their authorised applications. France was chosen for this case study due to the availability of PPP sales data at a higher spatial granularity. This has allowed generation of pesticide use maps that are closer to the reality and thus support the development of the resident exposure indicator.

\section{Materials and methods}

In this section, we first depict the data used followed by the methodological approach  (section \ref{sec:method_data}) combining various datasets to model pesticide applications on crops and the level of residents' exposure. Finally, the evaluation of the spatialisation with independent data source is described.

\subsection{Data}
\label{sec:method_data}
Data on pesticides are challenging to acquire, but some countries are more proactive in making such data available. France is one of the countries that releases datasets to the public. 
For this analysis, valuable datasets, summarized in \ref{tab:dataset}, were obtained in three main categories: Pesticides (\ref{sec:data_pesticide}), Agriculture (\ref{sec:data_agriculture}) and Population (\ref{sec:data_population}). A detailed description of the data can be found in the Supplementary material.

\begin{table*}[ht]
\caption{List of used datasets with description and references are split into three categories: pesticide, agriculture and population.}
\label{tab:dataset}
\begin{tabular}{|ll|p{8cm}|p{3cm}|}
\hline
\multicolumn{2}{l}{Dataset} & \textbf{Description} & \textbf{Reference(s)} \\ \hline
\multicolumn{2}{l}{\textbf{Pesticides}  (section \ref{sec:data_pesticide})} & & \\
 & Pesticide Sales - BNV-d & Database containing pesticide sales data in France from 2013 to 2021, based on declarations by registered distributors. & \cite{French-Government, postalcodes} \\
 & \ac{PPP} dataset & Digitalized data on plant protection products, fertilizers, and more from the French government. & \citep{Ramalanjaona,Cherrier} \\
 & \ac{PPDB} & Comprehensive database including pesticide chemical identity, physicochemical, human health, and eco-toxicological data. & \cite{PPDB} \\
 & EUPDB & EU database with details on approved and non-approved substances used in plant protection products. & \cite{EUPD} \\
 & Open EFSA & Platform for accessing information related to the European Food Safety Authority's (EFSA) scientific work. & \cite{OpEFSA} \\
 & LUCAS-Soil Survey & The ‘Land Use/Cover Area frame statistical Survey Soil’ (LUCAS Soil) is an extensive and regular topsoil survey that is carried out across the European Union to derive policy-relevant statistics on the effect of land management on soil characteristics. & \cite{lucas} \\
 & \ac{TFI} & The Pesticide Treatment Frequency Index created by the Solagro Group. & \cite{IFT} \\
\multicolumn{2}{l}{\textbf{Agriculture }(section \ref{sec:data_agriculture})} &  &  \\ 
 & \textit{Registre Parcellaire Graphique} (RPG) & Geo-dataset of agricultural parcels and crop type. & \cite{IGN} \\
\multicolumn{2}{l}{\textbf{Population} (section \ref{sec:data_population})} &  &  \\
 & Gridded Population Data & Data consisting of gridded information at a 200-meter resolution. Includes variables related to individuals' age distribution, household characteristics, and income. Derived from tax files. & \citep{populationLayer,filosofi} \\ \hline
\end{tabular}
\end{table*}

\subsubsection{Data interaction and schema}
\begin{figure*}
    \centering
    \includegraphics[width=0.70\textwidth]{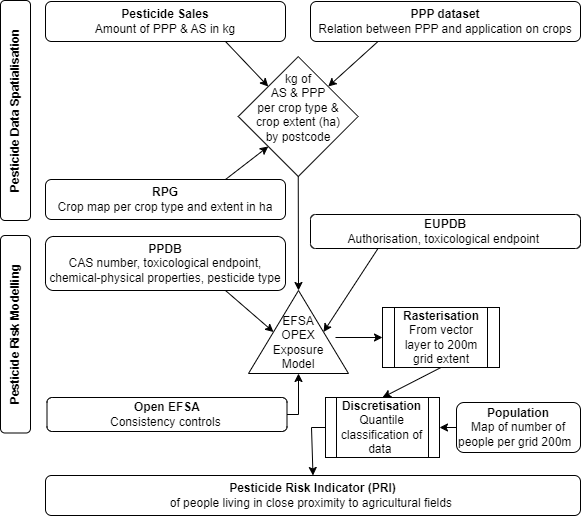}
    \caption{Overview of data utilisation in designing the Pesticide Risk Indicator (PRI): the diagram illustrates the specific segments of various datasets employed and their respective applications in the pesticide risk indicator's development.}
    \label{fig:draw3}
\end{figure*}

The data flow is presented in \ref{tab:dataset} and it is further schematized in \ref{fig:draw3}. 

\subsection{Methods}
\label{sec:method_method}

\begin{figure*}
    \centering
    \includegraphics[width=0.90\textwidth]{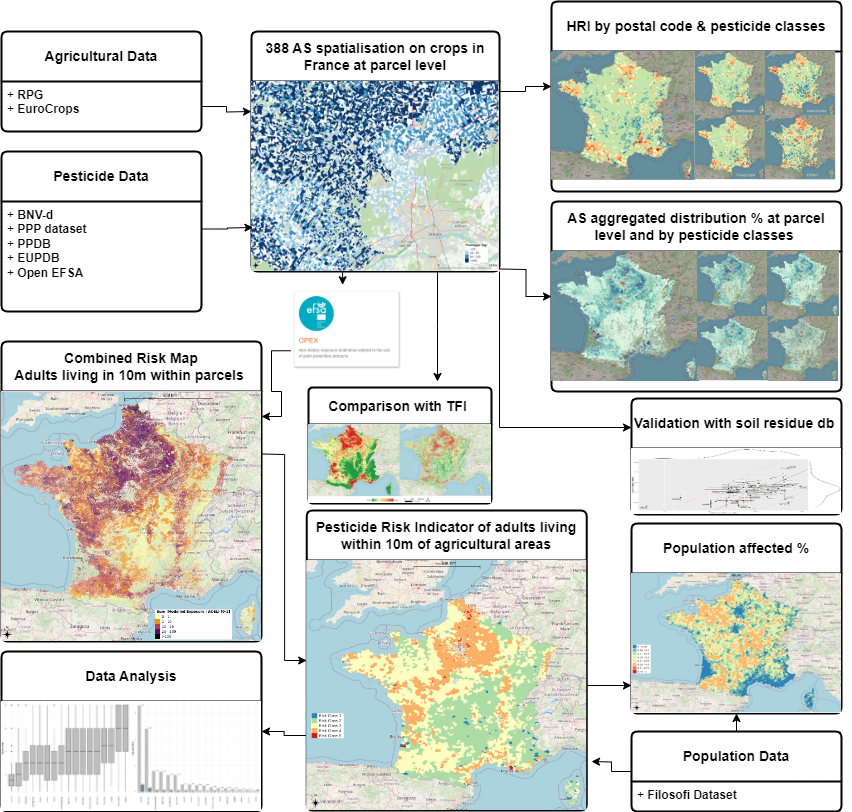}
    \caption{Methodological diagram for the development of the final indicator, including intermediate results and comparative assessment.}
    \label{fig:draw2}
\end{figure*}

\ref{fig:draw2} illustrates the full workflow of the methodology to calculate parcel pesticide loads and pesticide exposure. This includes the development of the \ac{PRI} indicator as well as intermediate results and the comparative assessment. The method was deployed for year 2018.

\subsubsection{Data Spatialisation}
The rationale behind this analysis is straightforward. Pesticide sales data in France were used as a proxy for actual pesticide use on crops in a given agricultural year (2018). Pesticide sales data in France is available at the postcode level, representing a territorial unit finer than NUTS3 (i.e., representing 6051 postcodes at country level). By combining this information with the authorisations of pesticides on crops (PPP dataset) and the parcel-level agricultural land use data (RPG), we estimated the potential quantity of pesticides applied on each single parcel. The harmonised crop classification from the EuroCrops dataset \citep{schneider2023eurocrops} was employed to ensure semantic legend consistency of the crop type classes. 
The pesticide sales dataset was curated, excluding entries with missing or zero postal codes and quantities. The result of this first data setup, resulted in 1.5 million records after cleaning (i.e., -9\% of the whole initial dataset). 
Authorisation of crops for each Active Substance (AS) was integrated, yielding 43 million records. This association was made through expert knowledge, linking the information on labels all to the respective crop types or categories. It is important to note that the crops reported on pesticide labels can vary in specificity. Some labels may specify detailed crops like artichokes, while others may use more general terms like cereals or arable crops. During the matching process, all possible cases were considered (e.g., cereals = soft wheat, durum wheat, maize, winter barley…). However, nearly 75\% of hypotetical pesticide applications did not align with actual crop coverage due to varied crop specificity on pesticide labels and those applications were not considered in the spatialisation of pesticide amounts on crops. Quantities of ASs and products were split across agricultural parcels based on postcodes, weighted in total area. The distribution did not take into account the potential application doses reported in the PPP dataset \citep{Ramalanjaona}. Instead, the fraction of each crop area was calculated relative to the total agricultural area of the postal code for each PPP and AS content. This fraction value was then multiplied by the quantities of ASs and PPPs to obtain a relative pesticide load expressed in kilograms. The application dose was subsequently calculated by dividing the resulting quantities by the area of the potentially affected crops by the use of PPPs.

The Pesticide Load (\(\text{PL}_{ijk}\)) can be calculated with equation \ref{eq:pesticide-load} where $i$ is a given active substance, $j$ is the postal code and $k$ the crop considered. The $PL$ is obtained by multiplying the quantity (kg) of a given Active Substance, $AS_{i}$ in a given postcode (\(j\)) by the proportion of authorised crop (\(k\)) for that active substance within the postcode:

\begin{equation}
{PL}_{ijk} = \text{AS}_{ij} \times \left( \frac{\text{authorised Crop}_{ijk}}{\sum_{k=1}^{n}\text{authorised Crop}_{ijk}} \right)
\label{eq:pesticide-load}
\end{equation}

All procedures were automated using a combination of R, PostgreSQL, and QGIS \citep{R,RStudio, PostegreSQL, QGIS} on the Big Data Analytics Platform of the JRC \citep{soille2018versatile}. The R scripts are available as well as the spatial data at postcode level. The link to have access to the data can be found in Supplementary material \ref{sec:link}.

\subsubsection{Pesticide Exposure and Risk Assessment}
This spatialisation procedure served to provide the pesticide application rates for the calculation of the modelled exposure and consequently the risk. 
Exposure is defined as the concentration or amount of an agent that reaches a target organism, system, or population within a specific frequency and duration  \citep{world2009principles, EFSA}. It is characterized through exposure scenarios which describe the circumstances of exposure, including sources, pathways, and amounts involved, and the organisms or populations exposed \citep{van2007risk}. These exposure scenarios are used to build exposure models that employ algorithms and equations to quantitatively estimate exposure doses via oral, dermal, or inhalation routes. The modelled values are compared to a reference value known as the Acceptable Operator Exposure Level (AOEL). The AOEL represents the maximum amount of AS to which an operator (or resident) may be exposed without experiencing adverse health effects, expressed in milligrams of AS per kilogram of body weight per day. The ratio between the estimated value from the models and the AOEL generates a Risk Quotient. A Risk Quotient below one indicates an acceptable risk, while values above one require further refined assessment to determine whether risk management measures need to be taken.

We relied on the "Guidance on the assessment of exposure of operators, workers, residents and bystanders in the risk assessment of plant protection products" released by EFSA in 2022 \citep{EFSA}. Specifically, our focus is on residents, defined as individuals who reside, work, attend school, or are present in an institution near an area that has been treated with pesticides. Their presence is unrelated to pesticide-related work, but their location may expose them to pesticides for up to 24 hours per day.

The exposure calculation was conducted for the following exposure routes: spray drift, vapour, surface deposit, entry into treated fields. The $50^{th}$ percentile of exposure was chosen for the entire assessment as suggested by the EFSA Guidelines. The model from EFSA was adapted to transform the information from a tabular format to a spatially explicit one. For each AS analysed in terms of pesticide load distribution based on their authorisations for different crops, all necessary data for exposure calculation were retrieved from different sources (PPDB, Open EFSA, EUPDB), and for each AS analysed, an exposure assessment was performed. In 2022, EFSA introduced an online tool called OPEX \citep{EFSA} for assessing non-dietary pesticide exposure, which includes the underlying equations used in the model. We used the online tool to validate our results. Following the guidance, we summed Hazard Quotients of individual AS to create a Hazard Index. This is a worst-case approach, from exposure point of view, assuming simultaneous application of all AS on a parcel, but also from a toxicological point of view, since no consideration of the modes of action and related possible contribution to combined effects was applied \ref{sec:expo}.   

Finally, the risk mapping was performed using a parcel level map, with calculations of the risk set at a 10-meter distance from the agricultural fields. The 10-meter distance was determined to be the most suitable, in accordance with the EFSA OPEX Guidance and in conjunction with the spatial resolution of the agricultural and population layers employed for developing the indicator. This assessment was carried out separately for both adults, assuming a body weight of 60 kg, and children, with an estimated body weight of 10 kg, representing the age range of 1 to 3 years old. The combined risk map is the results of the sum of four exposure routes:
\begin{enumerate}
    \item - Vapour exposure: it occurs when pesticides volatilize into the air and individuals inhale the vapour. This can happen during and after the application of pesticides.
    \item - Spray Drift: it involves the movement of pesticide droplets away from the target area during application. Drift can occur due to wind or improper application techniques.
    \item - Entry into Treated Fields: this route of exposure occurs when individuals enter areas recently treated with pesticides. It involves direct contact with treated surfaces, soil, or plants.
    \item - Dermal Transfer: it involves the direct contact of pesticides with the skin. This can occur during the application of pesticides or through contact with surfaces, tools, or clothing that have been contaminated with pesticides.
\end{enumerate}

\subsubsection{Towards the indicator of risk from non-dietary exposure}
In line with what has been shown previously, the combined risk map was integrated with the population distribution map. To conduct this analysis, the combined risk map was aggregated from a parcel-level resolution to match the grid size of the population map, i.e., 200 meters. Only overlaying pixels from both maps were considered to focus on residential areas close to agricultural fields. To enhance the clarity of the ultimate map produced, both maps underwent a reclassification process employing a scoring system that relied on the quantile distributions of values within each of the two maps. A scoring system ranging from 1 to 4 was applied to both the population distribution map and the combined risk map \ref{fig:indicator_calcultatio}. The scores of the different combinations were multiplied to result in an overall risk scoring and ranking in five classes. The five classes represent the increasing potential risk from 1 to 5. The two maps were multiplied together, resulting in a discrete indicator for mapping the risk to pesticides of residential population in agricultural areas. 
The resulting map can be presented in two different ways: either at the pixel level or by re-aggregating the indicator to the postcode boundaries, by average value.

\subsubsection{Population affected }
We also analysed the percentage of population potentially affected within the different indicator classes at postcode level. We performed this by dividing the number of people potentially exposed in our analysis with the total population living in the same area delimited by postcode boundaries. The resulting ratios were depicted into a map. It is to be noted that there is a discrepancy in the datasets utilized with the data on pesticide sales and crop mapping being from the year 2018, while the population distribution map is derived from the 2017 population census.

\subsubsection{Data aggregation}
AS aggregated distribution: to enhance the visualisation of areas and crops with potentially widespread pesticide use, we created aggregated distribution maps for ASs throughout France. We normalized the total spatialised ASs in each crop at the postcode level, scaling the dimensionless values from 0 to 1 for a comprehensive analysis. We made the decision to visually compare the Treatment Frequency Index (TFI, more can be found in \ref{sec:data_pesticide}) \citep{pingault2009produits, IFT, IFT2}, focusing on pesticide use intensity based on application rates and recommended doses, with the aggregated distribution map that normalizes total spatialised ASs in each crop at the postcode level. This comparative analysis aims to provide a comprehensive understanding of pesticide usage patterns by considering both intensity and spatial distribution across different agricultural areas (as described in section \ref{sec:Evaluation_of_spatialisation}). 
 
\subsection{Validation of the spatialisation } 
\label{sec:Evaluation_of_spatialisation}
To assess our results, efforts were made to utilize accessible data on residues of ASs in the soil. For this purpose, we used a pesticide soil residue database from the LUCAS in-situ survey \citep{orgiazzi2022lucas, lucas, vieirapesticides}. LUCAS Soil surveys 2018 for France has 683 locations sampled with pesticide residues in Soil [mg AS/kg of soil] covering 118 AS and pesticide metabolites, see sampling point map in  \ref{fig:LUCASsamples}.
The latitude and longitude coordinates of the selected soil French sites were used to extract corresponding pesticide dataset information from LUCAS Soil dataset. At the parcel level, our spatialised layer provided information on application rates or quantities of ASs. To convert these application rates into soil residues, the guidelines of the FOCUS (FOrum for the Co-ordination of pesticide fate models and their Use) Work group on Soil Modelling was followed, specifically using the first-tier equation that assumes worst-case concentrations immediately after application (t0)  \citep{boesten1997soil, EFSAPECsoil}.

The equation used is as follows:
\[
\text{{Initial PECsoil}} = \frac{{\text{{Application Rate}} \times (1 - \text{{Foliar Interception}})}}{{100 \times \text{{soil depth}} \times \text{{Bulk density}}}}
\]

For this analysis, the foliar interception value was set to 0 as a worst-case assumption, the soil depth was set to 20 cm as sampled in the soil dataset, and the bulk density values reported in soil for each site were utilized.
The calculated values, expressed in mg AS/kg of soil, were compared with those measured residues from the pesticide soil dataset. It is important to note that the overlap of the two datasets resulted in 50 pesticides present in both datasets, which is a lower number compared to the original 118 substances in the soil database and 388 substances from our spatialisation. This is because the substances in the soil database include pesticide metabolites and ASs that are no longer approved, while our layer is based on sales data of substances that are approved or recently withdrawn. Additionally, the number of sites included in the analysis decreased from 683 to 519 due to the exclusion of sites not included in the RPG layer. A dataset was created matching the pesticide residues analysed in the same postal code and crops for both the datasets. 
For all the analysis, we are only considering AS residues belonging to the same crop and postal code according to the LUCAS site positions.

\section{Results}

This section presents the results including detailed mapping of ASs, evaluation of resident exposure risks, and quantification of population affected.

\subsection{Spatial indicator for residential pesticide exposure and risks}
\subsubsection{Mapping the active substances potential use at parcel level.}
\begin{figure}[ht]
    \centering
    \includegraphics[width=0.5\textwidth]{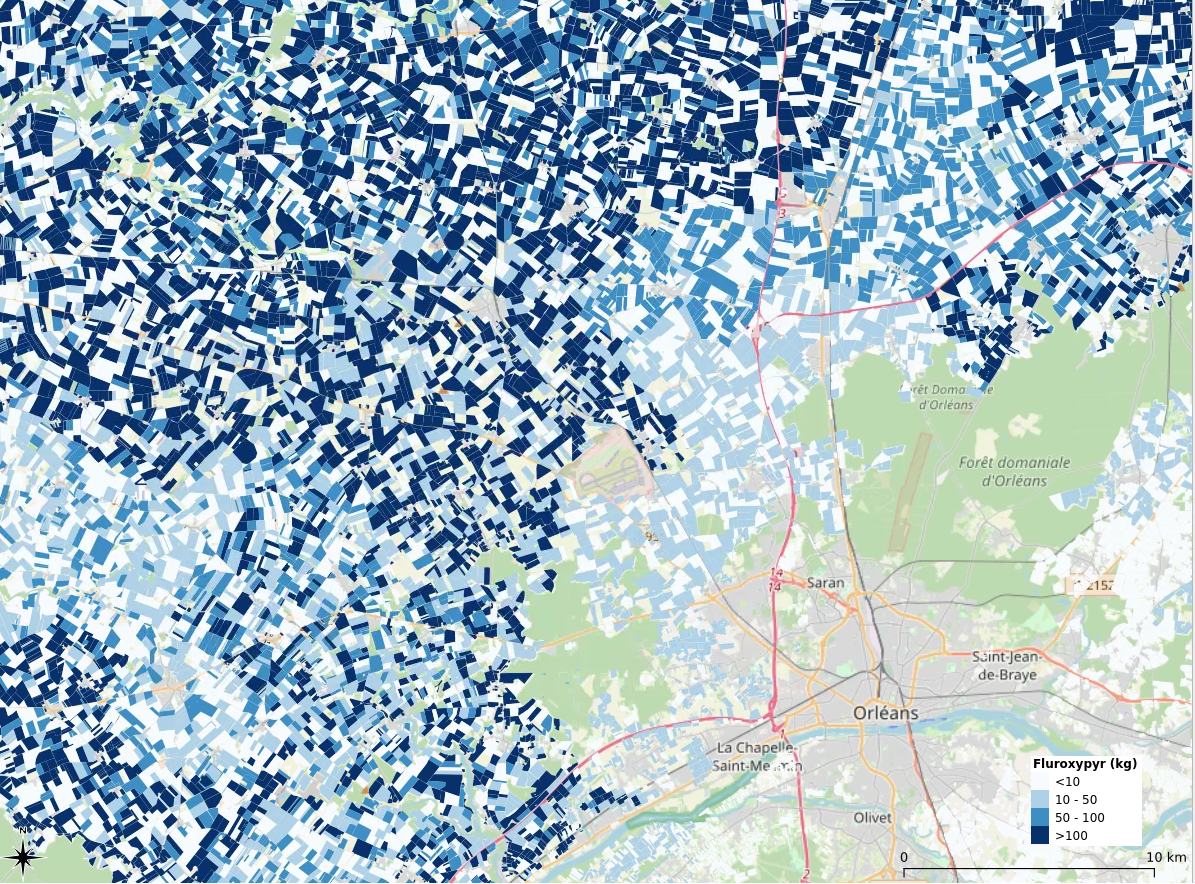}
    \caption{A highlight of the north-west Orléans agricultural area, where a potential application of the AS Fluroxypyr is estimated based on sales. The colour scale represents the potential applied amount in kg distributed among an area which is the areas belonging from per sum of the same crop type and areas within the same postcode. Globally, 388 AS and 9.5 M parcels were mapped }
    \label{fig:fluro}
    
\end{figure}

By consolidating and standardizing the quantities of all ASs, we created 388 maps (one for each AS), depicting potential pesticide loads at the parcel level in France, which comprises 9.5 million parcels. As an example, \ref{fig:fluro} displays a detailed map illustrating the potential application of the herbicide Fluroxypyr near the city of Orléans. Fluroxypyr is authorised for post-emergence use throughout the EU, specifically targeting broadleaf weeds and woody brush. It finds application in diverse settings, including cereal fields (e.g., wheat and maize), orchards, and vineyards. All these maps can represent the potential amount (in kg) of ASs applied on a single parcel, or the potential application rate, considering also the parcel extent (kg/ha). A list of all the AS analysed is provided in the Supplementary material \ref{tab:ASlist}.

\subsubsection{Fine scale resident pesticide exposure risk.}\label{sec:results_Resident_exposure} 
\begin{figure}[ht]
    \centering
    \includegraphics[width=0.5\textwidth]{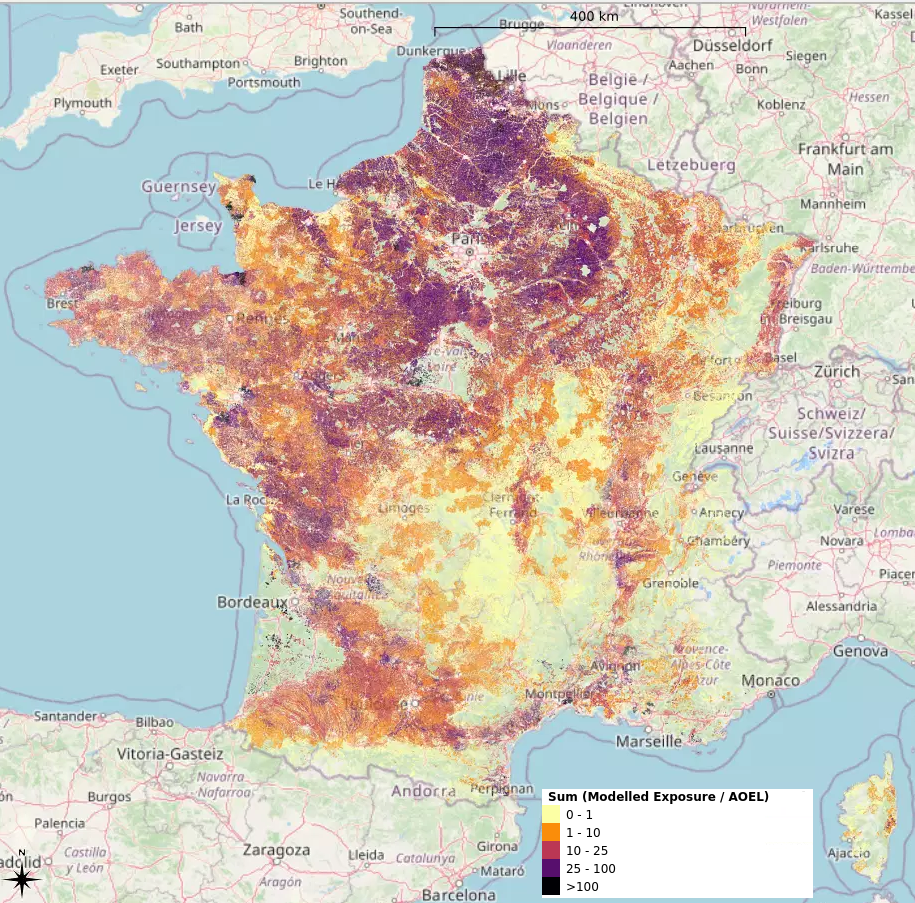}
    \caption{Potential combined risk of all the ASs, for adults living within 10 m from agricultural areas. The colour scale depicts these worst-case risk levels calculated as sum of risk quotients of the modelled exposure divided by the reference values AOEL. The risk calculation serves as an indicator of potential concern levels rather than providing exact risk values, as its primary purpose is to facilitate comparison of risks between different crops and regions rather than conducting an exact risk evaluation.}
    \label{fig:ExpAdults}
\end{figure}

After mapping out potential pesticide use, we produced a risk map measuring the combined exposure experienced by generic individuals living near agricultural fields. Here we are exploring the possible risk in these areas not yet considering if there are people living in that area. This potential combined risk of all the ASs to Adults within the 10-m buffer strips is shown in \ref{fig:ExpAdults}. As shown in the legend, the map depicts the sum of the ratio between EFSA modelled exposure and the toxicological endpoint AOEL. In \ref{fig:exas_children} is presented the combined risk map for children.

\subsubsection{Risk indicator applied to population.} 
\begin{figure}[ht]
    \centering
    \includegraphics[width=0.5\textwidth]{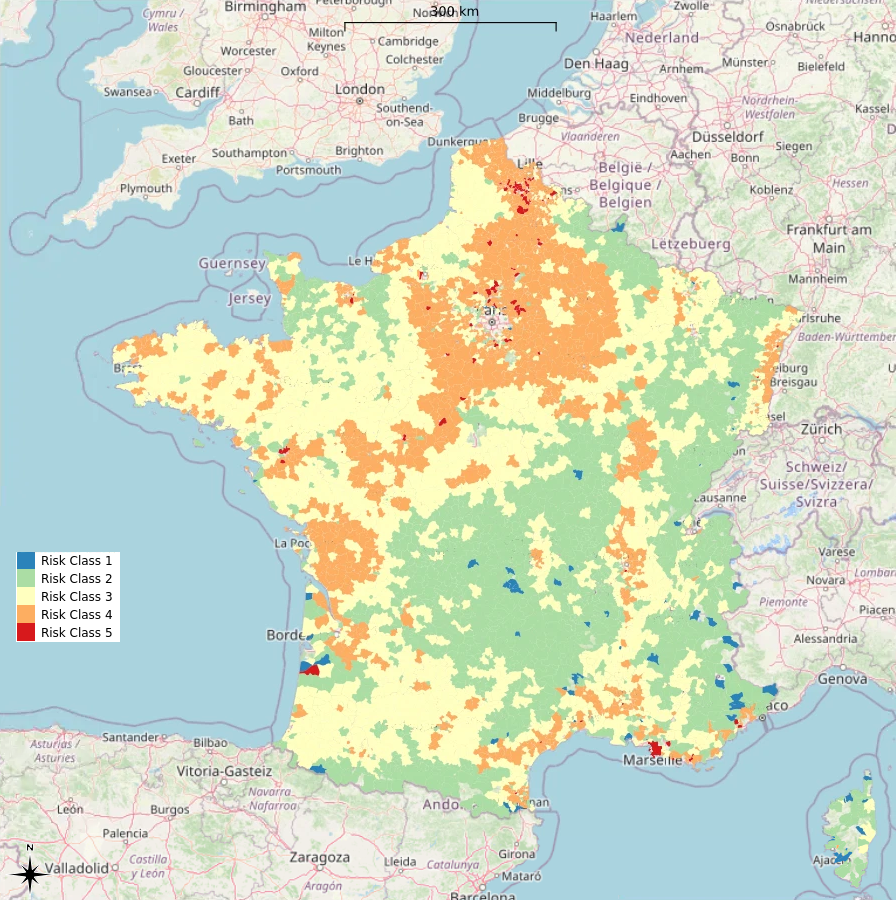}
    \caption{Pesticide risk indicator: integration of combined pesticide risk with French population distribution map at postal code level.  The results, initially detailed at a 200 square meter resolution, were aggregated by averaging pixel values and then displayed at the postal code level, providing a more comprehensive overview than the finer administrative division of municipalities in France. For the combined pesticide exposure assessment map, we performed a reclassification based on its quantile value distribution in conjunction with the Population distribution map. e.g. the higher the modelled exposure and risk near the field in combination with a higher number of people living in the relevant area, the higher the indicator value.}
    \label{fig:ExpAss}
\end{figure}

We then identified regions potentially affected by risk to the actual population by incorporating population maps. The results can be presented in two formats: a raster format with a pixel size of 200 x 200 meters and a vector layer delineated at the postcode level on the map (averaged values over postal code areas) as presented in \ref{fig:ExpAss}. We classified the indicator into five distinct classes ranging from low risk (class 1) to moderate and potentially elevated risk (class 5).  In \ref{fig:indicator_children} shows the indicator calculated for children.

The \ref{fig:routeofexposure} and \ref{fig:routeofexposure_ch} presented the four different exposure routes for adults and children respectively, showing that entry into the fields is a dominant route followed by vapour, while systemic and spray drift exposure are contributing less.  

\subsubsection{Potentially affected population}
\begin{figure}[ht]
    \centering
    \includegraphics[width=0.5\textwidth]{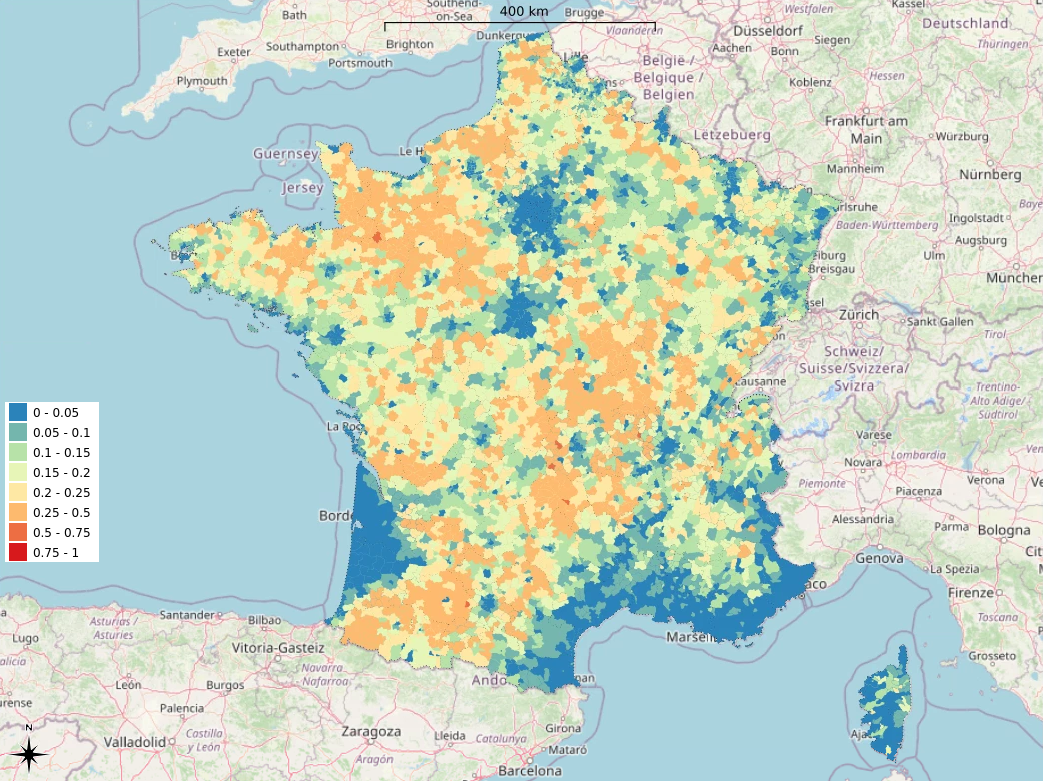}
    \caption{Exposure of the adult population to pesticides versus the total population at the postal code level.}
    \label{fig:Exppopper}
\end{figure}

To ascertain the proportion of France's population potentially exposed to pesticides, we conducted an analysis at the postal code level. The map in \ref{fig:Exppopper} shows the proportion of adult population exposed to pesticides per postcode. The proportion of exposed children is depicted in  \ref{fig:exposed_children}.   

\section{Validation of the spatialisation}
\subsection{Soil concentration comparison}
An initial analysis focused on comparing the crop types reported in both datasets. In the soil dataset, the crop type is determined in-situ, while in our spatialisation, the main crop is extracted from RPG records. Out of the total sites (n=520), in 93\% (n=482) of the cases, the reported crop type matched. For 7\% (n=37) of the cases, although the crop types did not match, they were likely rotational crops associated with the main crop of our spatialisation. In only 0.2\% (n=1) of cases, the reported crop type does not match, and it is unlikely to be a rotational crop (e.g., vineyards vs. grassland). 
Next, our focus shifted towards analysing the values of residues in greater detail.  \ref{fig:Scat2} displays a scatter plot showcasing the pesticide residue values in soil for our modelled data compared to the soil dataset on average.
It can be observed that the majority of points are distributed within a two-order-of-magnitude window around the diagonal, with the exception of Chlorpyrifos, Carbendazim and Prosulfocarb.
The marginal distributions of the two data series show that the distribution curve of our data is much less pronounced and elongated along its horizontal axis compared to the distribution of pesticide residue values in the LUCAS soil survey, which shows a less homogeneous distribution and a more pronounced clustering around the centre.
To compare the two datasets we calculated some model performance parameters. The coefficient of determination (R2) for this point distribution has a value of 0.14, indicating a weak relationship between the variables. The value of root mean square error (RMSE) is 0.02, representing good predictions. The Pearson correlation coefficient is equal to 0.37, suggesting a low positive correlation between the variables. More on model performance evaluation can be found in the Supplementary material (\ref{sec:modelperf}).

The data assessment continued from two perspectives: active substances and crops. In the first part, the focus was on individual active substances detected in various crops. Consequently, a procedure was developed to generate scatter plots of both residue data for a single active substance in relation to one or more utilized crops and conversely a scatter plot procedure for a single crop with several active substances.
\ref{fig:statsAScrops} shows bar graphs with the three correlation indices calculated for all active substances considered (\cref{sub:1}) and all crops (\cref{sub:2}). Also, this in-depth analysis does not improve the resulting correlations between the datasets. The sporadically high correlations reflect on low statistical power due to limited amount of population involved in calculating the indexes.

\subsection{Treatment Frequency Index comparison}
We conducted an additional effort to compare and validate our data using the pesticide Treatment Frequency Index (TFI) from the ADONIS map provided by the Solagro. Essentially, the municipal pesticide TFI serves as an estimation of pesticide utilization levels within each municipality in France. This estimation is derived from factors such as the municipality's crop rotation, the farming practices employed (whether conventional or organic), and regional reference TFIs, which are established through statistical or data locally surveyed. The resulting value serves as both a reference point for local farmers to gauge their own farm TFI and as an indicator of the potential risks associated with water, air, and food contamination stemming from the use of PPPs in agriculture. Through visual comparisons and analyses conducted at postal code level, we devised a strategy to crosscheck the TFI obtained from the \citep{pingault2009produits, IFT, IFT2} with the Aggregated Distribution map of AS.
This approach enabled us to evaluate the coherence and congruence between the TFI data and our results and could additionally offer a more nuanced understanding of pesticide use \ref{fig:IFT}.

\section{Discussion} 
In this work we developed an indicator to monitor the pesticide exposure of people living near agricultural fields.

\subsection{Interpretation of Results}
\subsubsection{Spatialisation of active substances sales data}
 The spatialisation of AS quantities and PPP usage on crops was a critical initial step in understanding how farmers applied the products they purchased in their respective regions. However, the sales data inherently came with a significant level of uncertainty. Some purchased products might be stockpiled for future use. Postcode information was also anonymized when it could potentially identify a small number of farmers, a privacy measure. Additionally, a product authorized for multiple uses might be used exclusively on one crop, contrary to our assumptions.
All reported pesticide sales within an administrative postal code are entirely utilized within the boundaries of that specific postal code, as we did in our work. This assumption, however, has a weakness. For instance, if a farm possesses fields spanning across three different administrative postal codes, and its main office is located in a fourth postal code, all pesticides acquired by that farm will be attributed to the fourth postal code. In reality, this postal code might not have received any pesticide treatment, yielding possible anomalies in the results. \cite{martin2023modelling} partially solved this issue with non-anonymised RPG data. These data cannot be shared due to privacy constrains.
A shortcoming of the RPG layer is that it does not cover all the agricultural parcels in France. We considered the option of enhancing the RPG layer by integrating missing information through the supplementary RPG complété dataset provided by the French government \citep{French-Governmentc}. This integration had the potential to significantly expand the coverage of parcels throughout France. However, unlike the conventional RPG dataset, this additional layer lacked a harmonised classification system of crop type. Aligning the two datasets would have been a time-intensive process, posing challenges in ensuring consistency and uniformity between the datasets. As a result, despite the potential increase in coverage, the absence of a harmonised classification made this approach less feasible for our analysis. We assess the percentage of crop missing by not considering RPG \textit{complété} as a complemental map to the RPG and we focused on the Département La Haute-Vienne (NUTS code FRI23). By comparing all RPG \textit{complété} parcels with the RPG ones, included in this administrative area, the land loss is 10\%. But if looking at only intensive agricultural crops and permanent crops, the percentage loss is 0.5\%. We conclude from this analysis that the enhancing from the RPG could be a marginal improvement.

 \subsubsection{Assessing possible pesticide exposure and risk for residents near agricultural fields} \label{sec:expo}
Our approach to estimating exposure and risks to people living close to agricultural fields represents a worst-case scenario, both in terms of exposure estimation and toxicological considerations by assuming simultaneous application of all ASs on a parcel without considering the modes of action and potential combined effects. It is important to note that the EFSA OPEX guidance also acknowledges weaknesses in constructing scenarios for residents due to a lack of available data. EFSA aims to initiate data collection efforts to address these gaps in the near future. 
In establishing the model, we considered dilutions for dermal absorption in the case of spray drift in a generic manner, following an expert judgment framework rather than considering the actual label-specific dilutions, which would have involved nearly 2000 different PPPs. Granules formulations were not taken into account. Furthermore, the timing of application was not factored in, and in the combined exposure assessment, it was assumed that all products were used simultaneously on a hypothetical treatment day for the agricultural parcel. It is important to recognize that this represents a simplification and a worst-case scenario. The risk values of the map (\ref{fig:ExpAdults}) ranges from 0 to greater than 100.
We need to be cautious in interpretation of these results and possible true concerns considering the worst-case assumptions detailed above. Also important to keep in mind is that this part of our work is looking into the theoretical risk to people in the vicinity of agricultural fields while this does not yet factor in whether there are people residing in these areas which follows in the next step. Values shown here should be only seen as a basis for comparing different regions and crops regarding risk ranges to residents. 
Upon analysing the map, it becomes evident that in the northern part of France, including Amiens, Rouen, Orléans, Troyes, and Reims, as well as in Bretagne and key wine-producing areas such as Alsace, Champagne, Burgundy, the Loire Valley, Bordeaux, and Provence, the potential risk could be elevated. Meanwhile, the risk falls into intermediate ranges almost everywhere else in France, except for the lower central part where the modelled risk falls within low risk levels for residents. The higher risk situations are likely related to outliers or peaks in sales data in specific postcodes. For example, the modelled exposure and the population affected indicate low pesticide risk in the city of Marseille, yet the risk map classified this area in the upper category of risk. Such situation might have been caused by higher concentration of agricultural holdings in this city and low numbers of crop parcels.
 If concentrating on single exposure routes (\ref{fig:routeofexposure} and \ref{fig:routeofexposure_ch}), it emerges that dermal transfer and spray drift are exposure routes of lower concern for resident, according to our potential pesticide applications. Entry into treated fields and vapour seem to be the ones driving the risk.  

 \subsubsection{Pesticide risk indicator map}
 The next step then is to include the actual population distribution. The resulting map (\ref{fig:ExpAss}) appears to be similar to the combined pesticide risk map. Some differences between the two maps, derive from the population distribution input, but also from the applied transformations to the pesticide risk indicator map: vector to grid data, discretisation and quantile classification, and along with data averaging at postcode level. The map values range from 0 to 16, as a result of the multiplied scores, and values are grouped into 5 risk classes. It is worth noting that in the lower risk ranges [class 1 and 2], there may be instances where the estimated exposure and resulting combined risk is elevated, but the population residing in those areas is very small. Conversely, the more elevated estimated risk classes show regions where factors such as landscape configuration, population density, and pesticide exposure assessment converge, which could lead to higher levels of pesticide exposure for the residents. The overall pattern is similar to the combined pesticide risk map identifying mostly the same areas to be investigated further as before. 
To gain deeper insights, we integrated the crop layer with the indicator map to perform an analysis of the crops involved in our study and assess which crops are most commonly cultivated near residential areas within agricultural sites. We accomplished this by creating a radius buffer of 100 meters from the centroids of each indicator pixel and examining the composition of the buffer zone in terms of crop coverage versus the average value of the PRI indicator \ref{fig:boxcrops}.

In this figure, the extents of 17 different crops can be observed, which together represent over 98\% of the total analysed area. Additionally, the figure presents boxplots illustrating the distribution of the indicator values for each crop throughout France. Crops with an average PRI value greater than 8 (i.e. PRI risk class 4) like vineyards, spring barley, potatoes and beet roots can be considered as crops impacting the pesticide exposure the most. More data are reported in \ref{table:crops}.

\begin{table*}[ht]
\centering
\begin{threeparttable}
\begin{tabular}{
  l
  S[table-format=6.0, group-four-digits=true]
  S[table-format=1.1]
  S[table-format=1.0]
}
\toprule
             & { Million Hectares} & {\% total analyzed buffer area} & {\% in whole France} \\
\midrule
Beet roots           & 0.01     & 1      & 3      \\
Vineyards         & 0.05     & 3        & 8      \\
Spring barley & 0.01       & 1       & 3     \\
Potatoes           & 0.01        & 0.6       & 6     \\
\bottomrule
\end{tabular}
\caption{crop composition of the selected buffer and extent vs crop extent in France}
\label{table:crops}
\end{threeparttable}
\end{table*}

These findings provide valuable insights into the crops contributing significantly to pesticide exposure in proximity to residential areas within agricultural sites. The predominant route of exposure can vary depending on the pesticide, application method, and environmental conditions. In many cases, dermal exposure is considered a significant route, especially for those directly handling pesticides, which, in this study, is not the case. Operators, and workers are supposed to use personal protective equipment (PPE), whereas residents are unlikely to wear these kinds of protections. This is also why, entry into treated fields can be a significant route, especially for individuals living in close proximity to agricultural areas. Vapour exposure and spray drift are more likely to affect individuals in the immediate vicinity of application sites and they are routes of exposure directly dependent from the application methods. In addition, vapour exposure is influenced by climatic conditions and pesticide volatility. Spray drift is mainly influenced by wind speed and direction, droplet size, spray equipment and height of application \ref{fig:IFT}

 \subsubsection{Exposed population}
 On average, the proportion of people potentially exposed due to living in the proximity of agricultural fields compared to the entire French population is 13\%. The map (\ref{fig:Exppopper}) shows that the percentage of population potentially exposed to pesticides is quite low when looking at larger towns like Paris, Rennes, Lyon, Nantes, Toulouse and so on. This depends on the density of the population but also on the proximity to agricultural fields. Then, some natural parks are depicted too with low exposed population as well as the Mediterranean coastal area of France. In \ref{table:popexposedperc} the exposed population is presented according to the indicator risk classes.

\begin{table*}[ht]
\centering
\begin{threeparttable}
\begin{tabular}{
  l
  S[table-format=6.0, group-four-digits=true]
  S[table-format=1.1]
  S[table-format=1.0]
}
\toprule
             & {\% average population exposed } & {N°postcode involved} \\
\midrule
Risk class 1           & 9     & 65     \\
Risk class 2     & 16     & 1734     \\
Risk class 3 & 15       & 2108      \\
Risk class 4        & 13        & 1311    \\
Risk class 5           & 9        & 86   \\
\bottomrule
\end{tabular}
\caption{subdivision of exposed population by indicator risk classes}
\label{table:popexposedperc}

\end{threeparttable}
\end{table*}

\subsection{Validation of the approach}
\subsubsection{Validation with soil residue estimates}. 
The same timing considerations seen before, can be applied to our attempt to compare the data with the soil residue database. In our work, pesticide residues in soil were calculated using the application rate resulting from the model we developed, applying the formula for calculating the PEC soil at time 0. This approach does not reflect the actual presence of pesticide residues in the soil. Additionally, the LUCAS soil residue data sampling does not account for potential pesticide applications in selected agricultural fields; it focuses mainly on assessing the qualitative state of European soils. While it is a unique survey, it is more focused on the history of pesticide applications on specific agricultural plots. This results in a comparison of residue data obtained in different ways (modelled vs analysed in the soil dataset) and for different investigative purposes. Low correlation could also be a consequence to the result of deposit from spray drift from adjacent parcels.
However, this does not diminish the importance of conducting an analytical comparison. Observed soil residues tend to be higher than our conservative estimate (initial soil concentration). This is also what emerges from the comparison with water samples in \citet{pistocchi2023screening}, and hints at a general underestimation of actual pesticide use/presence in the environment. One possible reason is the persistence of pesticides in the environment, reflecting past uses. 
\subsubsection{AS aggregated distribution and TFI}
 The TFI map provides insights into the intensity of pesticide application practices. On the other hand, the aggregated distribution map provides a broader view of the spatial distribution of pesticides, encompassing multiple factors that contribute to overall pesticide presence. By comparing these metrics, researchers and policymakers can gain a more comprehensive understanding of pesticide usage trends, potentially identifying areas with high intensity of use as well as areas where specific types of pesticides are more prevalent. This comparative visual analysis could be valuable for informed decision-making in agricultural and environmental management. The two maps exhibit predominantly similar class distributions; however, the key disparity in visualisation arises from the differing scales: TFI is represented at the municipality level, whereas the Aggregated Distribution of AS is depicted at the crop level. Consequently, the colours denoting crops appear less dense compared to the municipalities represented in TFI, influencing the overall visual contrast between the two maps.

By concluding, the term "indicator" in its name signifies its role as a signal identifying agricultural areas in France where the combined use of various ASs could potentially expose the resident population to pesticide risks, sometimes at elevated levels. Profitable agricultural zones, like vineyards, motivate farmers to maximize the use of available land for farming, giving rise to distinct features. Ongoing research aims to characterize a small buffer zone around urbanized areas throughout Europe, assessing the extent of these areas, the types of crops involved, and the PPPs used. Combining this analysis with the results of our study will provide a model for upscaling findings from a single member state to encompass the entire European continent.

\subsection{Comparison with Previous Studies}
\label{sec:comparison_Litterature}

Upon reviewing recent scientific publications, it is clear that various methodologies have been used to spatialise pesticides on agricultural land. These methodologies can be categorized as follows:

- Extrapolation from official surveys: researchers in different countries have used data from official surveys of agricultural practices to estimate pesticide application rates. Examples include studies in Germany \cite{ropke2004drips}, England  \cite{brown2007does}, and Belgium \cite{habran2022mapping}.

- Global gridded maps: a global approach involved creating comprehensive gridded maps displaying crop-specific application rates for various crop classes worldwide \citep{maggi2019pest}.

- Use of sales data: researchers utilized disaggregated country-based data on pesticide use and finer spatial resolution sales data, providing valuable insights into regional pesticide application patterns  \citep{schriever2007mapping, richards2012surveying}.

- Fine-scale data: certain regions have benefited from mandatory reporting by farmers at very fine scales, allowing for detailed analysis of pesticide use in these areas \cite{martin2023modelling}.

- Remote sensing and satellite imagery: advanced techniques were used to estimate pesticide use by observing changes in crop conditions over time, providing real-time data for understanding spatial variations in pesticide application \citep{ward2000identifying}.

- Crowd-sourced data: crowd sourcing platforms and citizen science initiatives have played a role in collecting localized data from farmers and local communities about pesticide use, providing nuanced insights into regional pesticide application practices \citep{haklay2014crowdsourced}.

Our approach is based on the meticulous analysis of fine-scale pesticide sales data combined with crop authorisation information. This method was chosen due to the availability of detailed parcel-level datasets, providing valuable insights into pesticide usage patterns with precision. Additionally, the integration of a harmonised classification system ensures uniformity and consistency across diverse datasets and regions, enhancing the accuracy of our analysis and facilitating meaningful comparisons.

\subsection{Recommendations for future research}
\label{sec:recomend}
In the upcoming endeavors, we have outlined several key areas of focus:

\textbf{Update and enhance spatialisation analysis}: by updating the spatialisation analysis by incorporating pesticide application rates and the number of applications. While these data are digitally available, the lack of harmonisation poses a challenge. Harmonizing this information is crucial for creating a solid dataset. Integrating application rates and the number of applications will significantly enhance the accuracy of our spatialisation. Modelling approaches will be applied to allocate data to specific crops, ensuring precision in our analysis. In addition, the Commission Implementing Regulation (EU) 2023/564, will be applicable from January 2026 and will make electronic record keeping obligatory. The real use data instead of proxy data will considerably increase the accuracy of our analysis.

\textbf{Organic vs. conventional parcels}: a dataset on organic and conventional parcels exists \citep{organic}. Although it is essential to recognize that organic farming does not imply the absence of pesticides, by incorporating it into our analysis it would increase the accuracy of the results in terms of better define treated vs less treated fields, considering also which AS are approved for use in organic farming.

An \textbf{in-depth analysis on PPPs} and their potential applications at parcel level could improve our knowledge of most hazardous pesticides on particular crops as well as it could increase the ability to look for pesticide substitutes to smooth and mitigate the pesticide risk in highlighted areas.

\textbf{Trend analysis of exposure}: utilizing nearly a decade's worth of crop distribution data at the parcel level, along with corresponding pesticide sales data, the trends in exposure assessment from the early 2010s to the present can be explored. This historical analysis will provide valuable insights into how exposures have evolved over time. Population maps, although not complete, will complement this analysis.

\textbf{EU-wide upscaling}: upscaling this analysis to cover the entire European Union is within future researchers' reach. To achieve this, there is the need to a transition to a more harmonised layer, such as the EUCROPMAP available for 2018 and 2022 \citep{d2021parcel,JRC, eucropmap2022}. EUCROPMAP offers high-resolution data at 10x10m pixel width, encompassing a wide range of crops (19 crop types). Replacing the farmers’ declaration parcel dataset with EUCROPMAP would allow this case study in France to be extended to other MS where data are available, like Denmark and Northern Italy, ensuring a consistent approach across regions.

\textbf{Population data enhancement}: Currently relying on French population distribution data, future EU-wide analysis will leverage the Global Human Settlement Population Grid layer (GHS-POP) from the JRC  \citep{GHS-POP-R2023A}. GHSL provides global, comprehensive and high-resolution dataset \citep{JRCb}, enabling us to pinpoint resident populations in agricultural areas with accuracy. An experimental disaggregation of population on a grid with enhanced resolution for Europe, compared to the global product at 100 m resolution, will be attempted to improve the exposure accuracy.

Recent studies such as from \citet{galimberti2020estimating} and \citet{udias2023} helped to perform our analysis as precursors that tried to implement a spatialisation of pesticide uses at EU scale, on which this study is clearly an improvement enabled by previously inaccessible data. Other ongoing European projects, including the Horizon 2020 SPRINT project\footnote{\url{https://sprint-h2020.eu/}}, have the potential to enhance our analysis. This is possible because these projects share similar objectives, allowing for valuable data sharing and collaboration.

These steps represent a comprehensive strategy to enhance the depth and scope of our analysis, ensuring a robust and insightful assessment of pesticide usage patterns in agricultural landscapes at different levels.

\section{Conclusions}
This study introduces a robust indicator designed to assess the potential pesticide exposure risk for residents living near agricultural fields in France. By meticulously analysing and spatialising a diverse range of datasets, we have developed a detailed model to estimate potential pesticide loads at a parcel level. The detailed spatial results allow to derive a picture of how relevant non-dietary exposure might be in certain areas and to identify areas where a more in-depth analysis might be required. Considering that we use a combination of several worst-case assumptions with several associated uncertainties, the results for the potential risks should not be considered an accurate risk assessment but a first indication of levels of concern that can only be used to tailor follow up actions and to compare among crops and regions. While our comparative assessment with existing datasets such as the LUCAS soil survey and TFI revealed areas for improvement, it represents a significant step towards a more nuanced understanding of pesticide risk. This effort not only underscores the need for high-resolution, harmonised data in pesticide risk assessment but also provides a strong foundation for extending this methodology across the European Union. With ambitious pesticide reduction targets on the horizon, our indicator serves as a valuable tool to assist stakeholders and policymakers in steering towards a safer and more sustainable agricultural landscape.


\section*{Acknowledgements}
We gratefully acknowledge the support of this research by the JRC Exploratory Research program through the PESTIRISK project. We thank Solagro for providing the ADONIS IFT dataset.

\section{Declaration of generative AI and AI-assisted technologies in the writing process}
During the preparation of this work, the author(s) used GPT@JRC Prototype to increase the manuscript readability. After using this tool/service, the author(s) reviewed and edited the content as needed and take(s) full responsibility for the content of the publication.

\bibliography{sample.bib}

\appendix

\onecolumn

\clearpage

\setlength{\cftfignumwidth}{1.4cm}
\setlength{\cfttabnumwidth}{1.5cm}

\tableofcontents
\listoffigures
\listoftables

\clearpage

\setcounter{figure}{0}
\setcounter{table}{0}
\setcounter{page}{1}

\section*{Supplementary material}

\label{AppendixA}
\captionsetup{list=no}
\renewcommand{\thetable}{Supplementary Table S\arabic{table}}
\renewcommand{\thefigure}{Supplementary Fig. S\arabic{figure}}

\section{}
\subsection{harmonised Risk Indicator 1}
\label{sec:hri}
\begin{figure}
    \centering
    \includegraphics[width=1\textwidth]{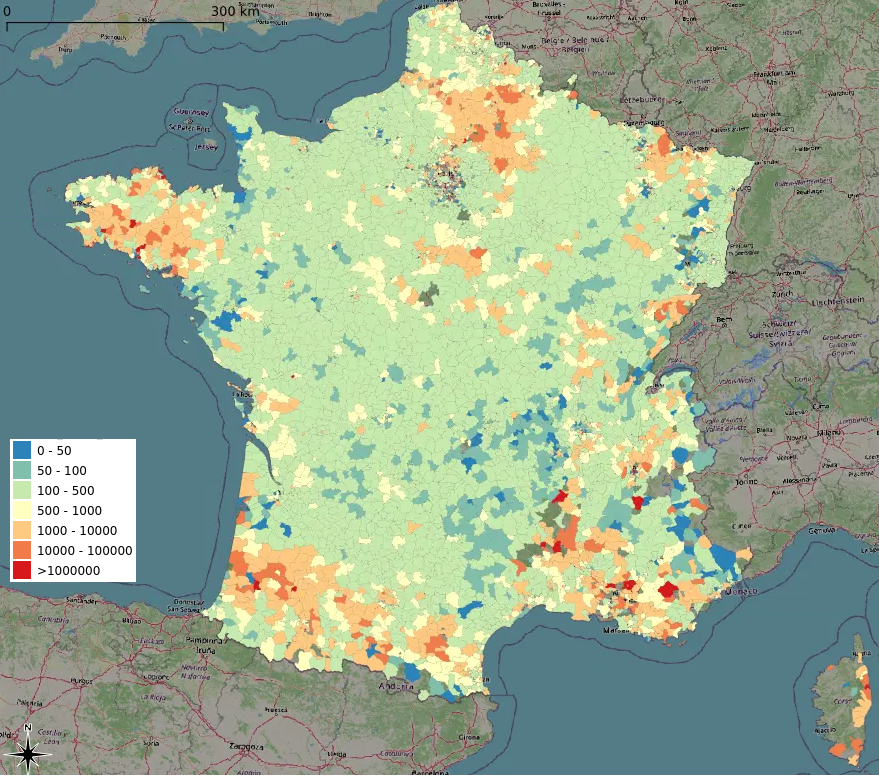}
    \caption{HRI1 by postcode: HRI1-2018 vs HRI1-2013}
    \label{fig:AS_HRI}
\end{figure}

\begin{figure}
    \centering
    \includegraphics[width=1\textwidth]{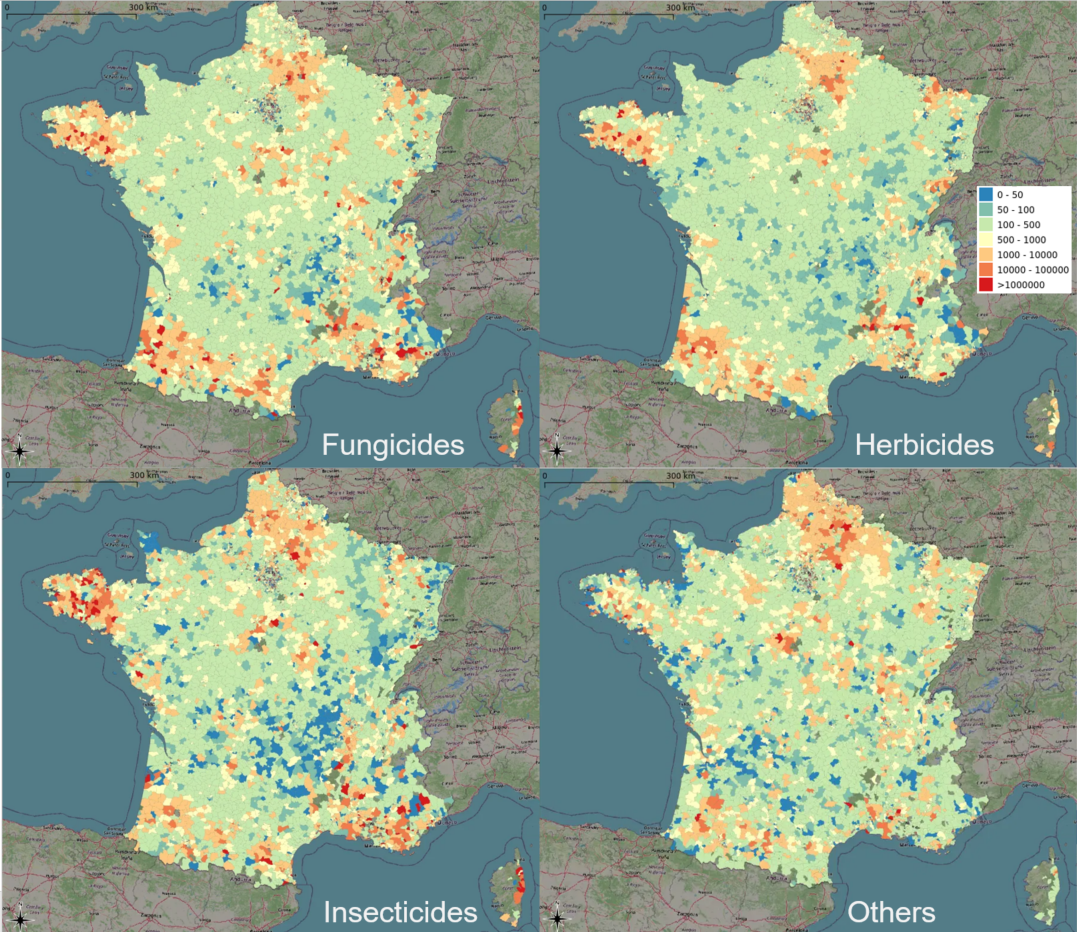}
    \caption{HRI1 by postcode: HRI1-2018 vs HRI1-2013 by group of pesticide}
    \label{fig:AS_HRIgroupped}
\end{figure}

\begin{figure}
    \centering
    \includegraphics[width=0.7\textwidth]{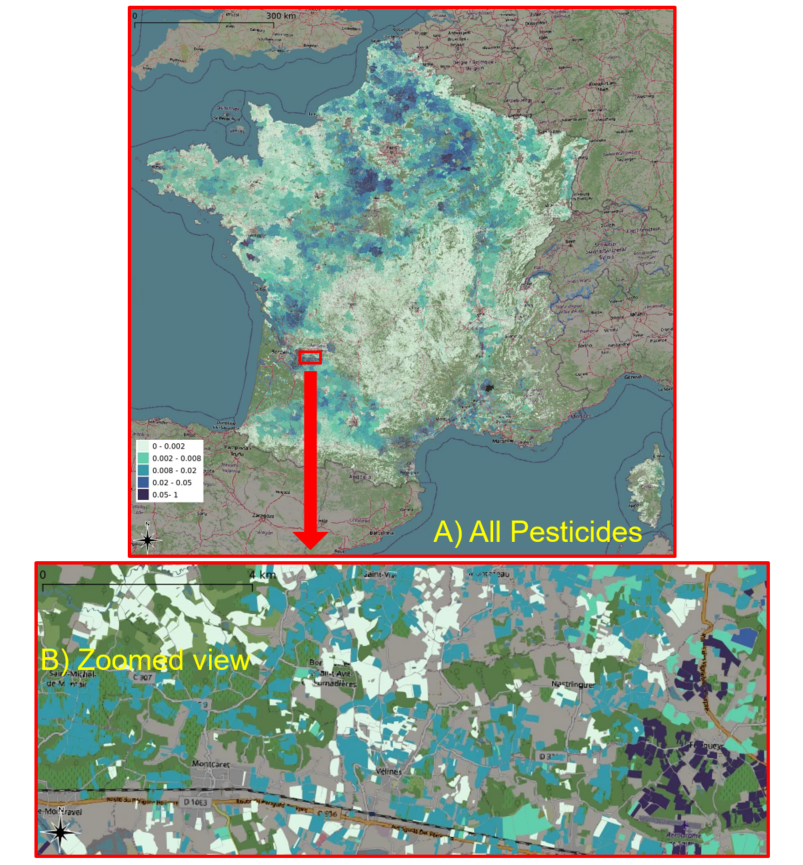}
    \caption{Aggregated Distribution of AS at parcel }
    \label{fig:ASaggrgated}
\end{figure}

\begin{figure}
    \centering
    \includegraphics[width=1\textwidth]{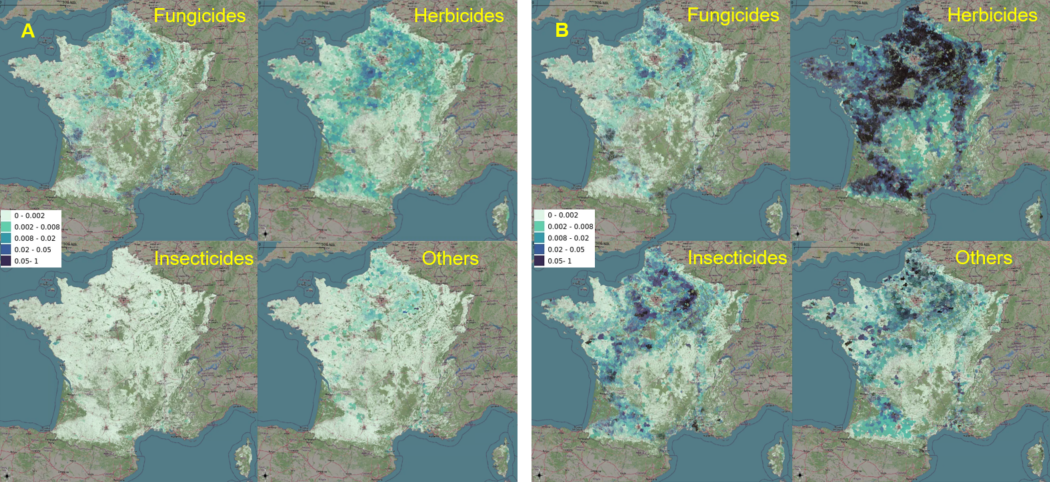}
    \caption{Aggregated Distribution of AS at parcel levels by group of pesticide: A) the aggregated distribution by ASs of the pesticide types considered together within the same range from 0 to 1. This visualisation allows to see differences in magnitude within the 4 selected pesticide types. B) Each pesticide type is considered singularly within the range 0 to 1. This visualisation shows the differences within each pesticide type.}
    \label{fig:AS_groupped}
\end{figure}

The harmonised Risk Indicator 1 (HRI1), is computed nationally using pesticide sales data, categorized by pesticide types and assigned specific weights. In France, pesticide sales data are available at the postcode level from 2013 to 2022. Therefore, we applied the same methodology used to spatialise the data for the year 2018, and calculate pesticide quantities for the year 2013. This approach enabled us to calculate HRI1 not only for 2013 over France but also at the postcode level. The data were weighted and categorized following the guidelines outlined in the Methodology for calculating the harmonised Risk Indicators annexes  \citep{eurostat21, eurostat21b}. Using the generated map and computing HRI1 for the available years, we were able to monitor the trend of HRI1 at a remarkably detailed scale over the past decade. 
To elaborate further, we utilized fine spatialisation techniques to calculate the HRI1 at a more precise scale. Additionally, we assessed the overall distribution of different AS in percentage terms. The HRI map, showcasing the calculated values for 2018 and 2013 (\ref{fig:AS_HRI}, \ref{fig:AS_HRIgroupped}), along with the corresponding map illustrating the percentage distribution of AS (\ref{fig:ASaggrgated}, \ref{fig:AS_groupped}). \newline
\ref{fig:AS_HRI}) was developed as an example of increasing granularity that can be reached with such data disaggregation. It appears that Bretagne and the area around Lille, as well as the southern areas of France are the most impacted by a change in the sales of pesticides. Our analysis uses just two points in time (2013 and 2018). It illustrates the value for looking at trends at a finer spatial scale, but to investigate trends more in depth would require some additional analysis over the years. \newline

\subsection{Dataset detailed description}
\subsubsection{Pesticides}
\label{sec:data_pesticide}

We utilized several datasets to gather information on pesticides:

- The \textbf{pesticide sales} dataset, obtained from the BNV-d (\textit{Banque Nationale des Ventes de produits phytopharmaceutiques par les Distributeurs agréés}) database, provided data on pesticide sales (Plant Protection Product – PPP and Active Substance - AS) in France from 2013 to 2021 \citep{French-Government}. This dataset was based on declarations made by registered distributors and aggregated by postal code \citep{postalcodes}. We used this dataset for taking into account the amount of \ac{PPP} and \ac{AS} by postal code.

- The \textbf{\ac{PPP} dataset} from the French government website offered digitalized data on plant protection products, including authorised and withdrawn products, product uses and doses, risk phrases, active substances, crop information, and parallel trade permits \citep{Ramalanjaona,Cherrier}. We mainly used this dataset to draft the relation between \ac{PPP} and authorisation on crops.

- The \textbf{Pesticide Properties Database (\ac{PPDB})} provided pesticide chemical identity, physicochemical, human health, and eco-toxicological data. We used this database to translate active substance names from French to English and collect toxicological endpoints for exposure assessment analysis \citep{PPDB}.

- The \textbf{EU Pesticide Database (EUPDB)} provided information on approved and non-approved active substances, including low-risk substitutes within the EU. It also included Maximum Residue Levels in food products and data on emergency authorisations of plant protection products granted by Member States \citep{EUPD}. We used this database for collecting \ac{AS} authorisations and toxicological endpoints.

- The \textbf{Open EFSA} platform served as a resource for accessing information related to the European Food Safety Authority's (EFSA) scientific work. It provided details on ongoing assessments, non-confidential versions of dossiers and studies, meeting agendas and minutes, and information about experts involved in the process \citep{OpEFSA}. We used the Open EFSA platform for consistency checks within different datasets.

- The \textbf{LUCAS Soil Survey 2018}, conducted by the European Commission, is a module of the Land Use/Cover Area frame Survey (LUCAS)  providing systematic sampling   and analysis of pesticide residues in soil across European territories \citep{d2020harmonised,lucas}. This survey provided insights into the environmental impact of agricultural practices and contributed to evidence-based decision-making and the formulation of agricultural policies \citep{orgiazzi2022lucas,vieirapesticides}.

- The pesticide \textbf{ Treatment Frequency Index (\ac{TFI})} from the ADONIS map\footnote{The ADONIS map was provided by SOLAGRO and is available on a viewer on \url{https://solagro.org/nos-domaines-d-intervention/agroecologie/carte-pesticides-adonis}.} estimated pesticide utilisation levels within each municipality or greater areas in France. It considered factors such as crop rotation, farming practices, and regional reference ac{TFI}s to gauge potential risks associated with water, air, and food contamination \citep{pingault2009produits, IFT, IFT2}.

\subsubsection{Agriculture}
\label{sec:data_agriculture}
We utilized the \textbf{\ac{RPG}} is a geo-referenced dataset of agricultural parcels in France \citep{IGN,French-Governmentc}. Each parcel is associated with declared cultivated crop types. The crop types were semantically harmonised into English EuroCrops classification \citep{schneider2023eurocrops}.

\subsubsection{Population}
\label{sec:data_population}
We utilized a gridded \textbf{population dataset} at a 200-meter resolution, which included variables related to age distribution, household characteristics, and income. The dataset was derived from tax files and adress data privacy by grouping cells with fewer households into larger aggregates (Filosofi - Income, poverty and standard of living in 2017 - gridded data system localized social and fiscal file) \citep{populationLayer,filosofi}. \newline

\subsection{Data access FTP}
\label{sec:link}
The data and R scripts related to the manuscript can be downloaded from the Joint Research Centre's Open Data Portal. The files include ExposedPOP\_AD\_postcode.gpkg, ExposedPOP\_CH\_postcode.gpkg, PRI\_AD\_postcode.gpkg, and PR\_CH\_postcode.gpkg, which contain detailed information on populations exposed to pesticides and Pesticide Risk Indicators (PRI) at the postcode level for specific regions. Additionally, a copyright.txt file provides the necessary copyright information. These resources are essential for researchers and policymakers interested in assessing pesticide exposure risks and formulating appropriate management strategies. For more details and to download the files, visit https://jeodpp.jrc.ec.europa.eu/ftp/jrc-opendata/DRLL/PESTIRISK/. \newline

\subsection{Model performance evaluation}
\label{sec:modelperf}
Model performance was evaluated using three commonly used statistical indicators:
\textbf{Pearson’s correlation coefficient} and \textbf{coefficient of determination (R2)} describe the degree of collinearity between simulated and measured data. The correlation coefficient is an index that is used to investigate the degree of linear relationship between observed and simulated data \citep{moriasi2015hydrologic}.

\textbf{The root mean square error (RMSE)} is the square root of mean square error (MSE). The MSE is also known as standard error of the estimate in regression analysis. The RMSE is measured in the same units as the model output response of  interest and is representative of the size of a typical error \citep{moriasi2015hydrologic}.\newline

\newpage
\subsection{Figures and Tables}

\begin{figure*}[ht]
    \includegraphics[width=18cm]{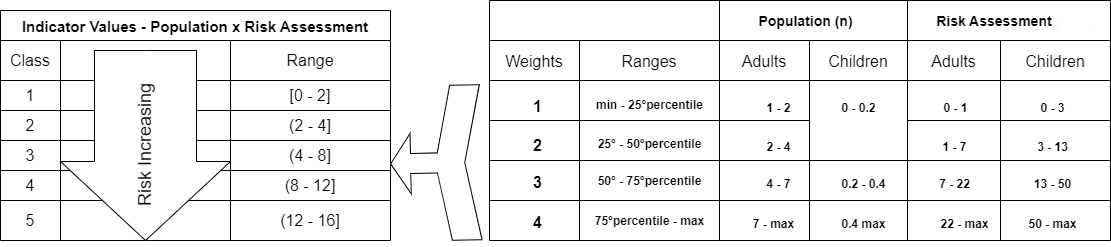}
    \caption{This figure represents the data flow from individual maps of population distribution and combined pesticide Risk to discretisation using a quantile classification. }
    \label{fig:indicator_calcultatio}
\end{figure*}

\begin{figure*}[ht]
    \centering
    \includegraphics[width=0.8\textwidth]{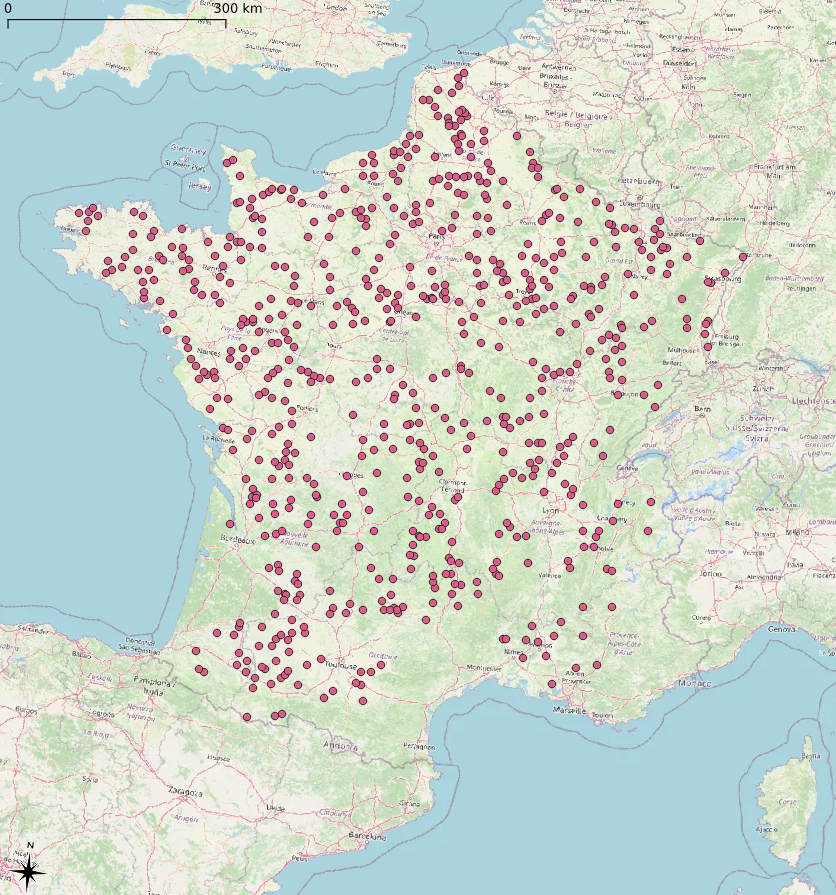}
    \caption{LUCAS Soil surveys (2018) provides 683 samples of pesticide residues in Soil [mgas/kgsoil] with 118 Active Substances and pesticide metabolites}
    \label{fig:LUCASsamples}
\end{figure*}

\begin{figure*}[ht]
    \includegraphics[width=18cm]{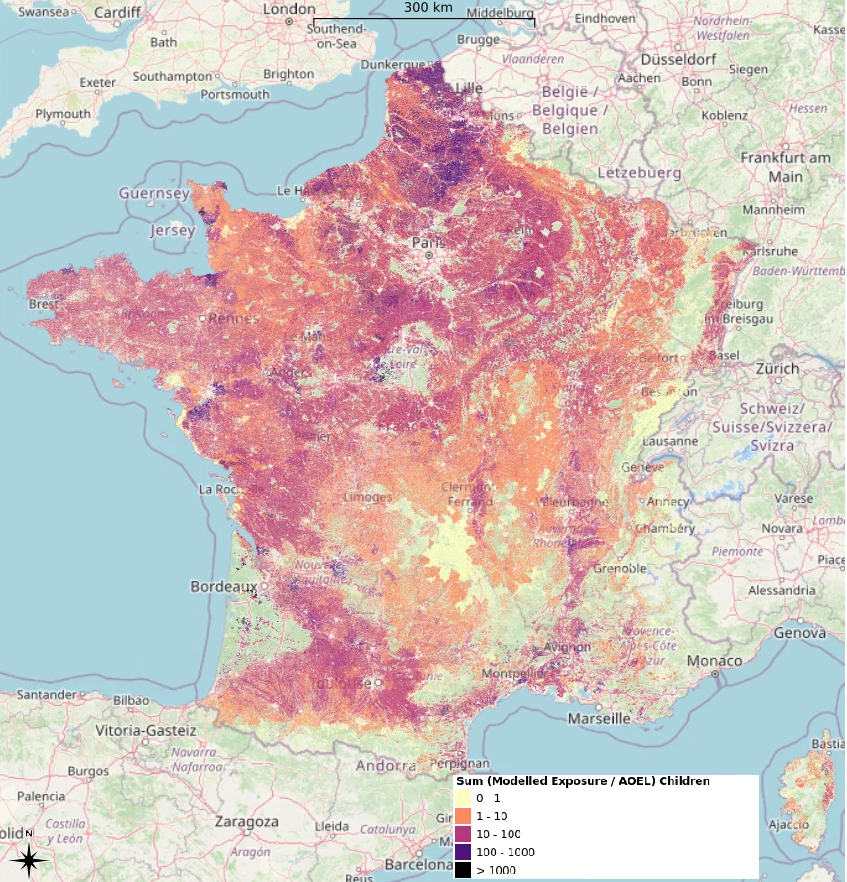}
    \caption{In this figure, it is presented the combined Risk assessment for Children - 10 meters from agricultural fields}
    \label{fig:exas_children}
\end{figure*}

\begin{figure*}[ht]
    \includegraphics[width=18cm]{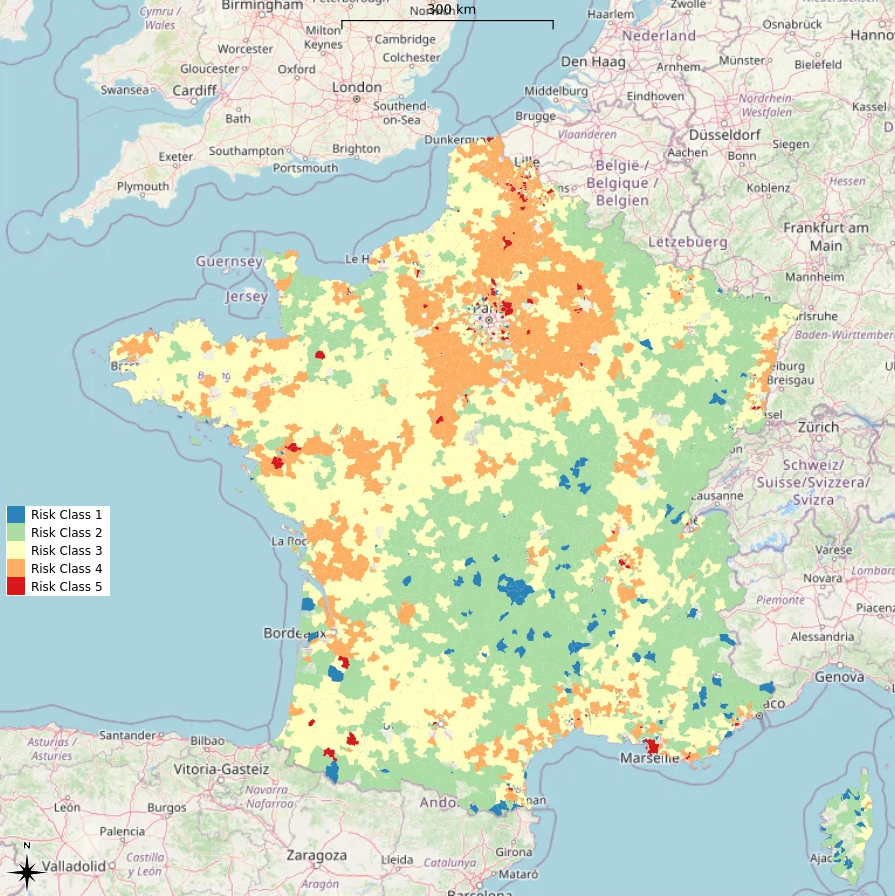}
    \caption{In this figure, it is presented the risk indicator map for Children, living in agricultural areas,  exposed to pesticides}
    \label{fig:indicator_children}
\end{figure*}

\begin{figure*}
    \centering
    \includegraphics[width=0.80\textwidth]{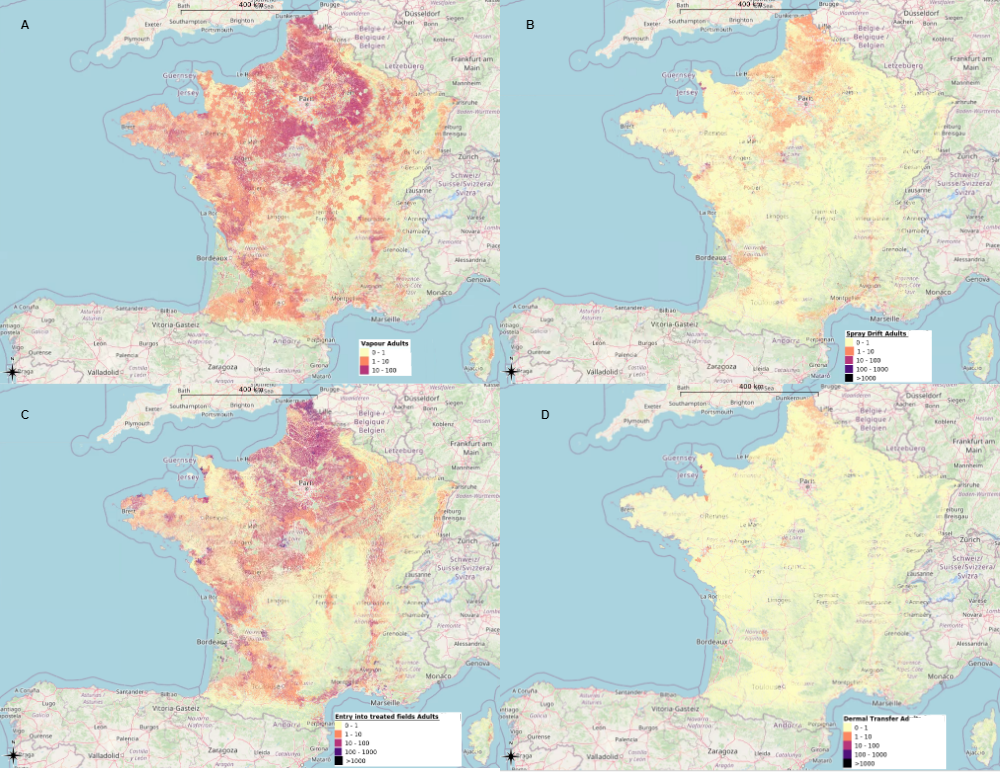}
    \caption{These are the four routes utilized by the model to assess the pesticide risk exposure for resident (adults). A) Vapour, B) Spray Drift, C) Entry into treated fields, D) Dermal transfer}
    \label{fig:routeofexposure}
\end{figure*}
\begin{figure*}
    \centering
    \includegraphics[width=0.80\textwidth]{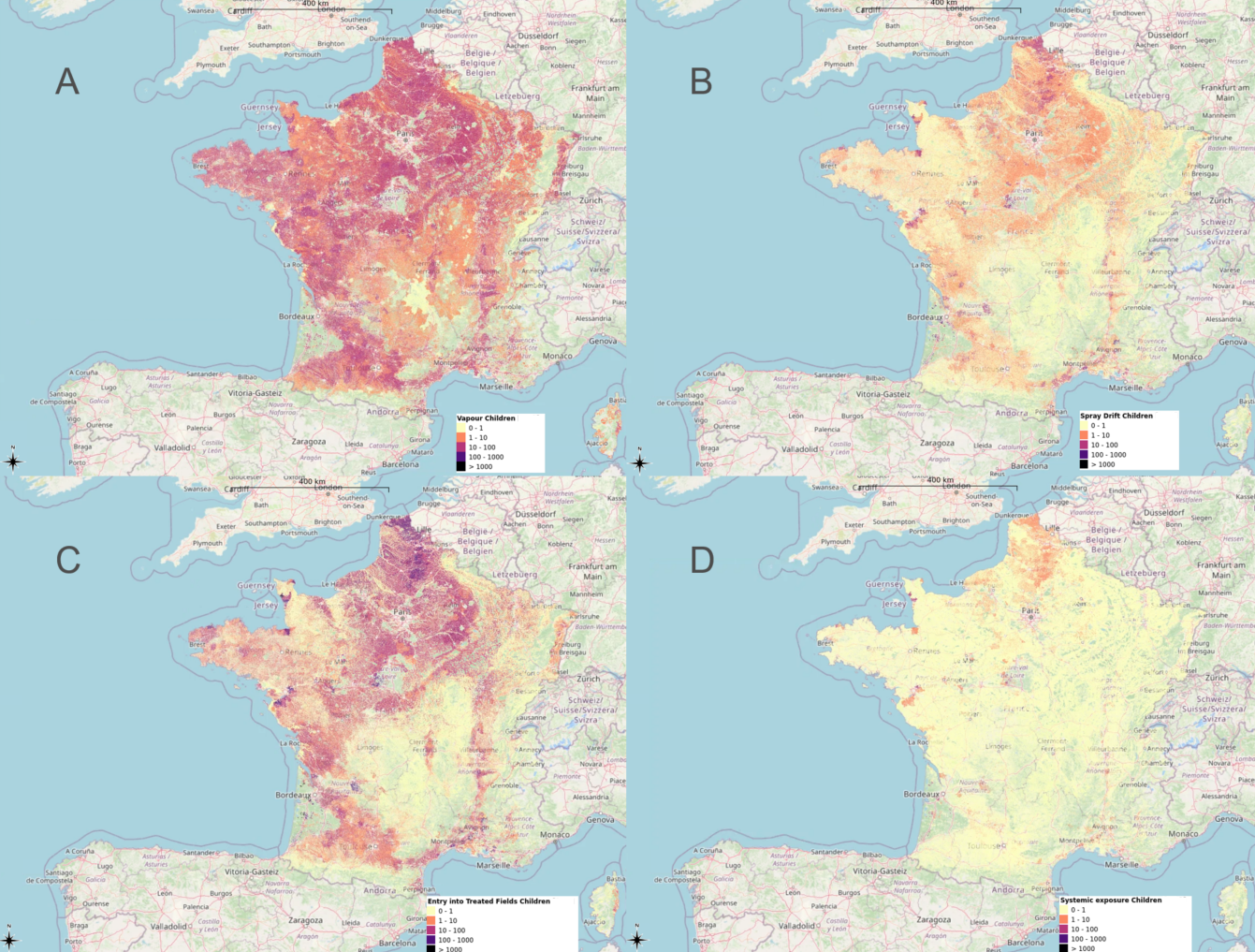}
    \caption{These are the four routes utilized by the model to assess the pesticide risk exposure for resident (children). A) Vapour, B) Spray Drift, C) Entry into treated fields, D) Systemic exposure}
    \label{fig:routeofexposure_ch}
\end{figure*}

\begin{figure*}[ht]
    \includegraphics[width=18cm]{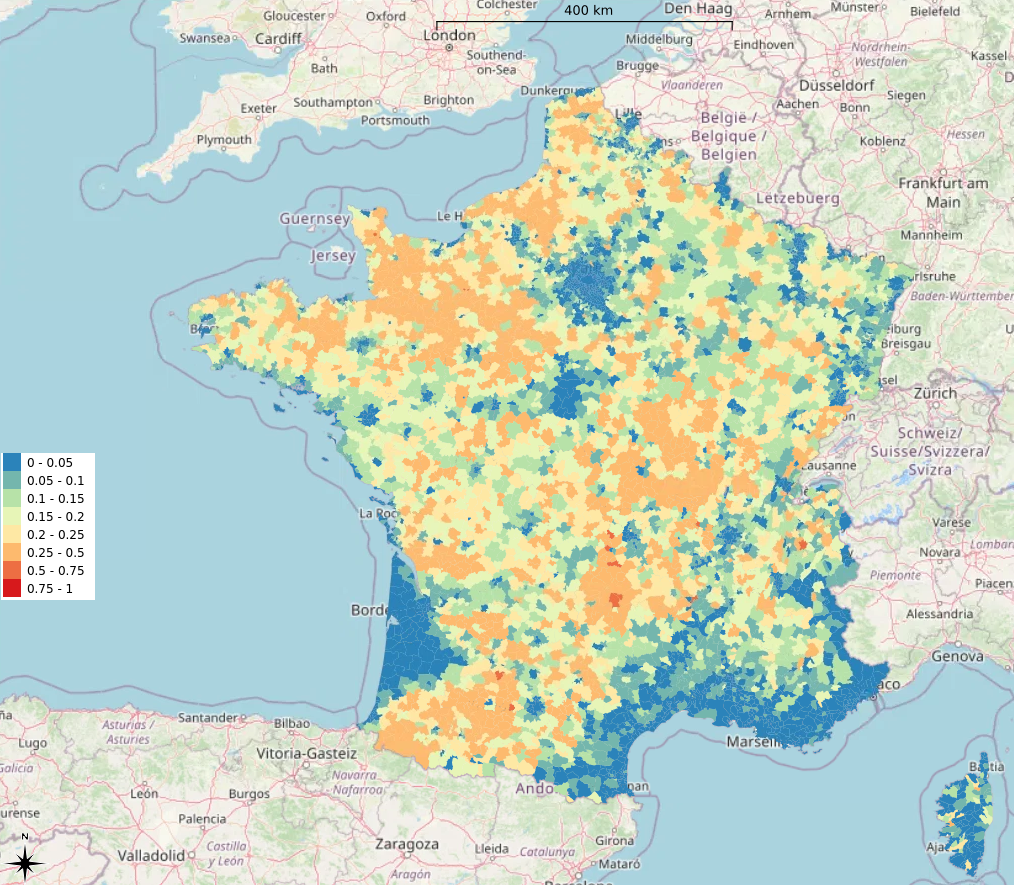}
    \caption{In this figure, it is presented the children exposed to pesticides i.e. the sum of children selected in the risk assessment per postcode, divided by all children in the same postcode}
    \label{fig:exposed_children}
\end{figure*}

\begin{figure*}
    \includegraphics[width=18cm]{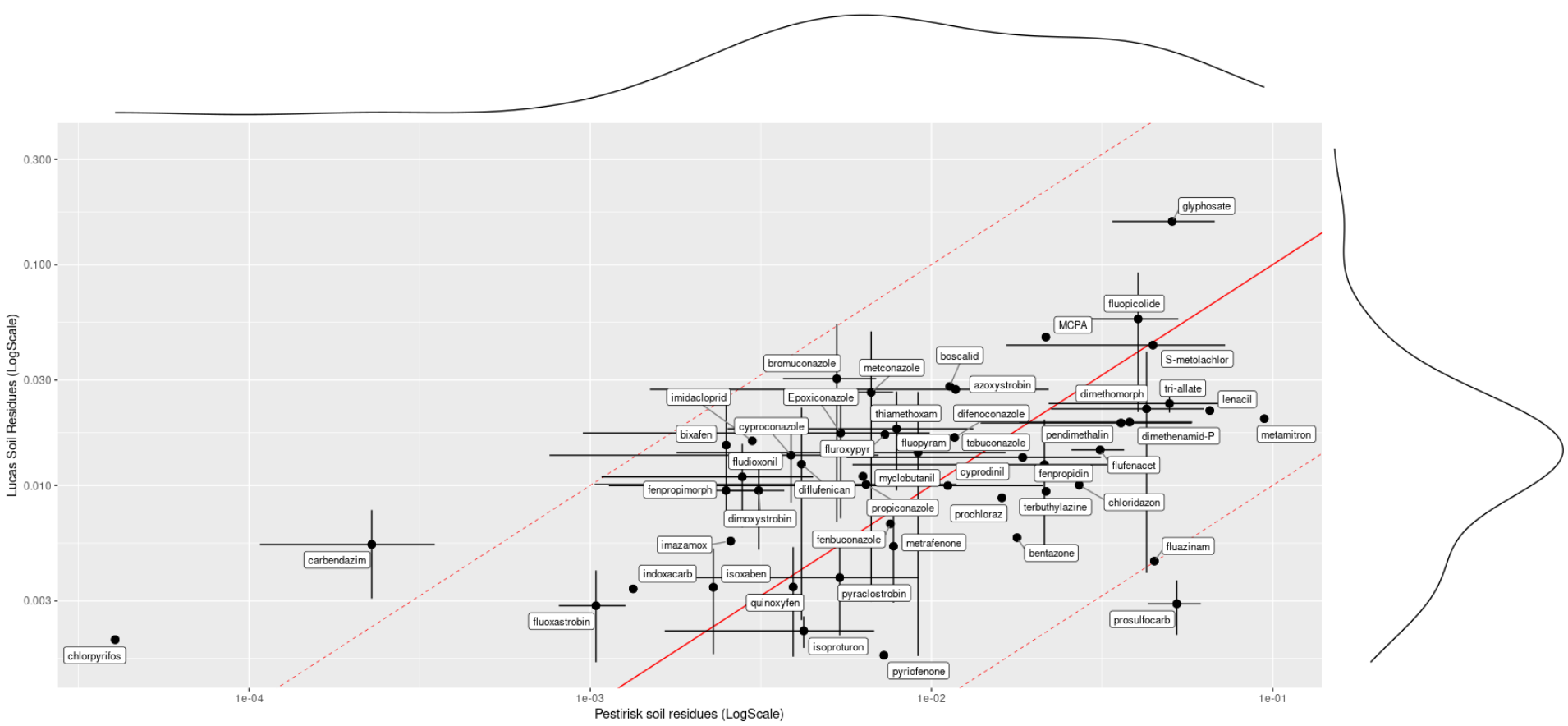}
    \caption{Scatter plot of the average modelled pesticide residue values in soils compared to the LUCAS soil database. The plot is in a logarithmic scale and includes standard deviation bars, marginal distribution, and a range that extends one order higher and lower than the diagonal (highlighted in red).}
    \label{fig:Scat2}
\end{figure*}

\begin{figure*}
\includegraphics[width=1\textwidth]{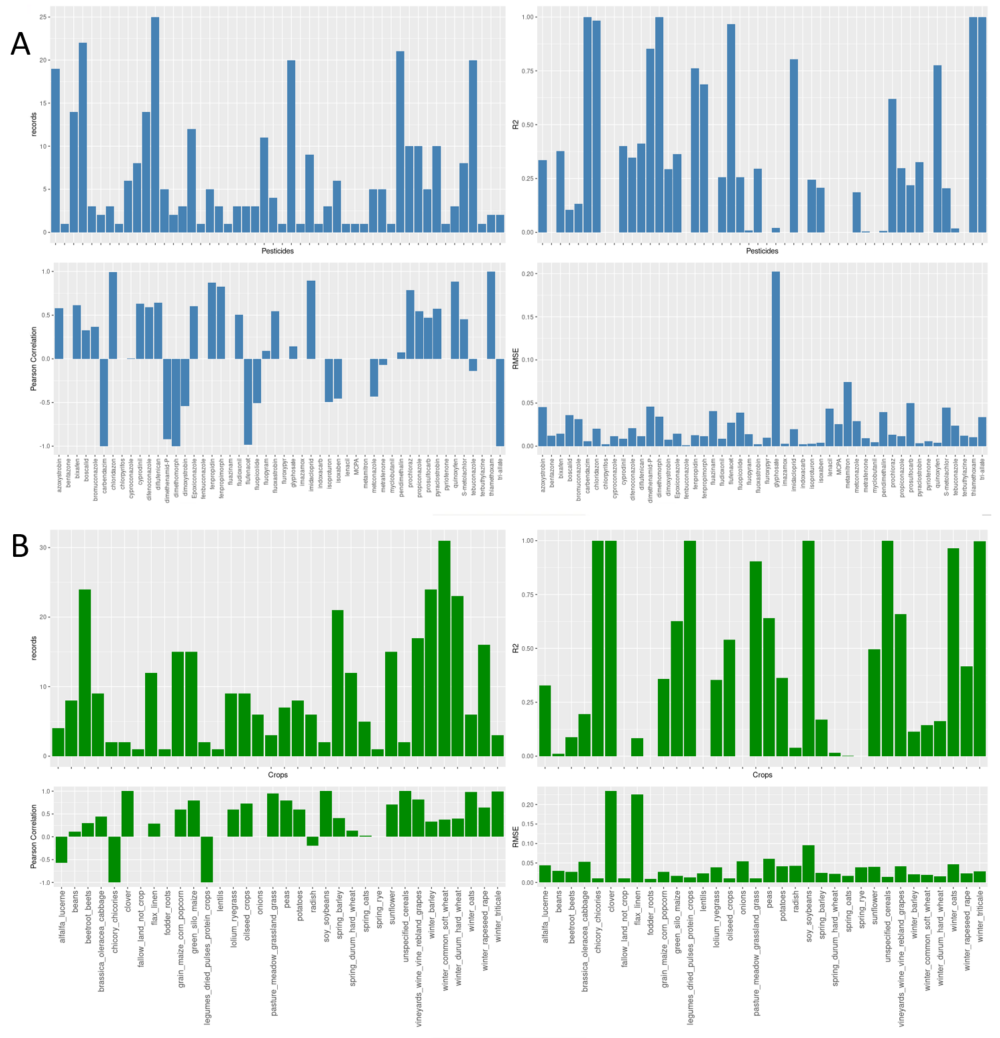}
       \begin{subfigure}{1sp}
       
       \refstepcounter{subfigure}\label{sub:1}
       \refstepcounter{subfigure}\label{sub:2}
       \end{subfigure}
\caption{Statistical analysis within scatter plots was considered with three different model performance parameters: number of records, R2, Pearson Correlation and RMSE for individual Active Substance (AS) across all crops (A) and for all AS together within each crop (B). (A) the AS were fixed and the scatterplots were based on crops were AS were found. (B) the Crops were fixed and the scatterplots were based on Active Substance distribution.}
\label{fig:statsAScrops}
\end{figure*}

\begin{figure*}[ht]
   \centering
    \includegraphics[width=1\textwidth]{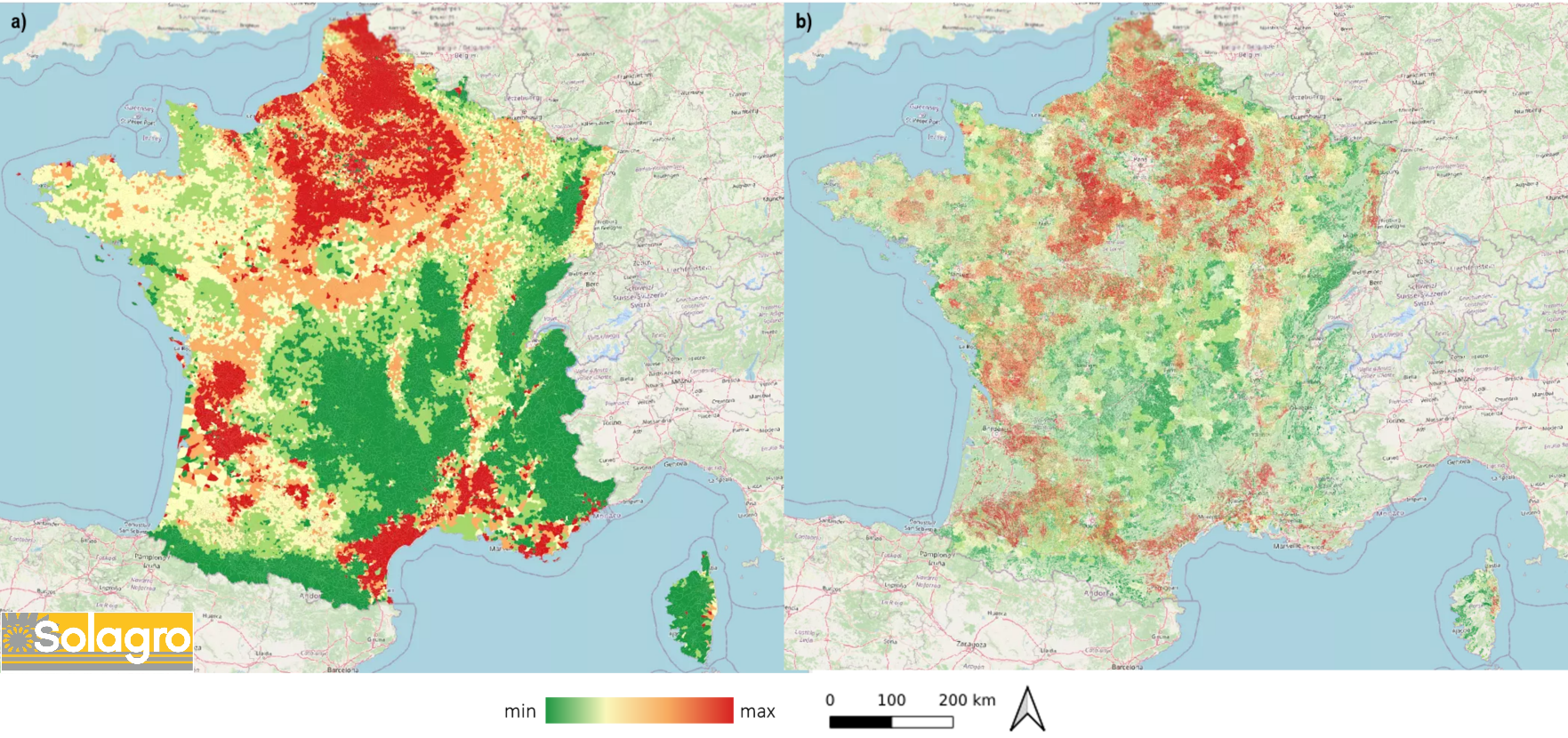}
    \caption{Comparison between Treatment Frequency Index (TFI, panel a) and Aggregated Distribution of Active Substance (AS) at parcel level (panel b). On the reader left hand side, the Adonis map is presented at the Municipality level scale (Source: Pesticide Adonis map, Solagro, 2022). On the right hand side, the Aggregated Distribution of AS at parcel level is outlined. Both maps have the same color schema from green to red. For TFI it goes from 0 to 21.9 whereas for the AS Aggregated distribution it goes from 0 to 1.}
    \label{fig:IFT}
    
\end{figure*}

\begin{figure*}[ht]
    \includegraphics[width=18cm]{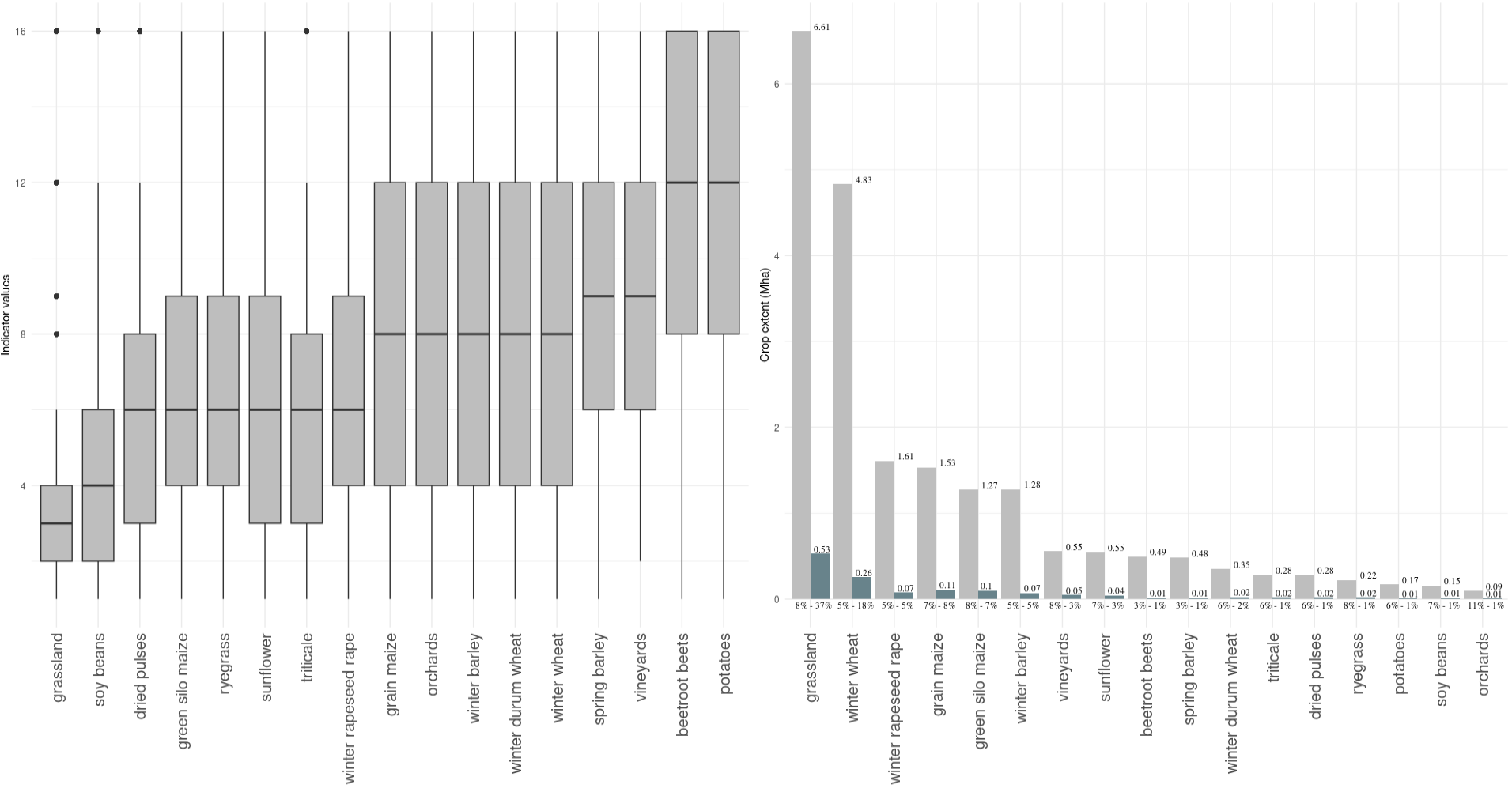}
    \caption{In the provided boxplots on the left, we present the distribution of indicator values within the 17 selected crops. On the right side of the figure, the bars, depict the distribution of crop coverage in France, measured in hectares. This representation allows us to visualize both the variation in indicator values across these 17 crops and their respective extents within the agricultural landscape of France. The two additional percentages at the bottom of the bars represent the proportion of crops extents in the selected areas in respect to the whole crop extent in France and the proportion of the composition all crop areas included in the selected areas (sum of all 100 meter buffer areas divided by crop type).}
    \label{fig:boxcrops}
\end{figure*}

\begin{figure*}[ht]
    \includegraphics[width=18cm]{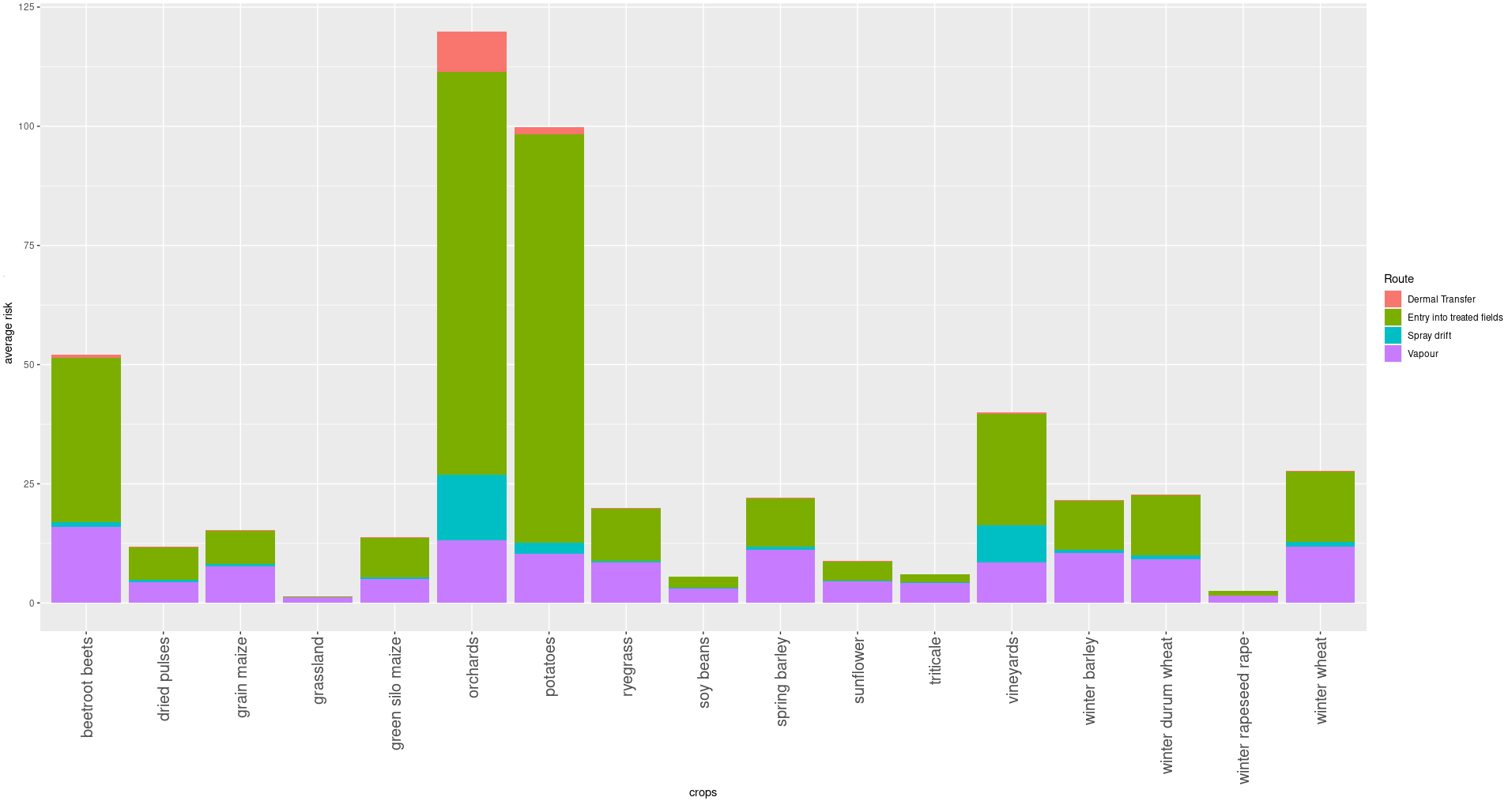}
    \caption{averaged pesticide exposure risk for selected crops coming from the buffer analysis and categorized by the four exposure routes}
    \label{fig:exproute_ind}
\end{figure*}

\clearpage

\subsection{List of Active Substances}
\begin{longtable}{ll}
Active                                                              & CAS-Number   \\
(E)-dodec-8-enyl acetate& 38363-29-0   \\
(E,Z)-octadeca-3,13-dienyl acetate                                  & 53120-26-6   \\
(Z)-11-tetradecenyl acetate                                         & 20711-10-8   \\
(Z)-dodec-8-enyl acetate                                            & 28079-04-1   \\
(Z)-dodec-9-enyl acetate                                            & 16974-11-1   \\
1,4-dimethylnaphthalene                                             & 571-58-4     \\
1,5-Diaminopentane                                                  & 462-94-2     \\
1-decanol                                                           & 112-30-1     \\
1-dodecanol                                                         & 112-53-8     \\
1-naphthylacetamide                                                 & 86-86-2      \\
1-naphthylacetic acid                                               & 86-87-3      \\
2,13-octadecadienyl acetate                                         & 86252-65-5   \\
2,4-D                                                               & 94-75-7      \\
2,4-DB                                                              & 94-82-6      \\
5-decenol                                                           & 56578-18-8   \\
5-decenyl acetate                                                   & 38421-90-8   \\
6-benzyladenine                                                     & 1214-39-7    \\
7,9-dodecadienyl acetate                                            & 54364-62-4   \\
8-Dodecenyl acetate                                                 & 37338-40-2   \\
8-dodecenol                                                         & 40642-40-8   \\
8-hydroxyquinoline                                                  & 148-24-3     \\
9-tetradecenyl acetate                                              & 16725-53-4   \\
Aureobasidium Pullulans                                             & 67891-88-7   \\
Bacillus thuringiensis Berliner subsp.   kurstaki                   & 68038-71-1   \\
Bordeaux mixture                                                    & 8011-63-0    \\
Decanoic acid                                                       & 334-48-5     \\
Diallyl trisulfide                                                  & 8008-99-9    \\
Dicamba-Dimethylammonium                                            & 2300-66-5    \\
Dipotassium Phosphite                                               & 13492-26-7   \\
Epoxiconazole                                                       & 133855-98-8  \\
FENOXYCARB                                                          & 72490-01-8   \\
Ferrous sulfate heptahydrate                                        & 7782-63-0    \\
Geraniol                                                            & 106-24-1     \\
Gibberellin A4 mixture with Gibberellin   A7                        & 8030-53-3    \\
Hexadecyl Acetate                                                   & 629-70-9     \\
Lecanicillium lecanii                                               & 67892-35-7   \\
MCPA                                                                & 94-74-6      \\
MCPB                                                                & 94-81-5      \\
Mineral oil, petroleum distillates,   hydrotreated light paraffinic & 64742-55-8   \\
Octanoic Acid                                                       & 124-07-2     \\
Polyalkyleneoxide modified   heptamethyltrisiloxane                 & 27306-78-1   \\
Potassium Laurate                                                   & 10124-65-9   \\
Propineb                                                            & 9016-72-2    \\
Rapeseed oil                                                        & 8002-13-9    \\
S-metolachlor                                                       & 87392-12-9   \\
White mineral oil                                                   & 9206-23-56   \\
abamectin                                                           & 71751-41-2   \\
acequinocyl                                                         & 57960-19-7   \\
acetamiprid                                                         & 135410-20-7  \\
acetic acid                                                         & 64-19-7      \\
acibenzolar-S-methyl                                                & 135158-54-2  \\
aclonifen                                                           & 74070-46-5   \\
acrinathrin                                                         & 101007-06-1  \\
adoxophyes orana granulovirus souche   bv-0001                      &              \\
alpha-cypermethrin                                                  & 67375-30-8   \\
aluminium phosphide                                                 & 20859-73-8   \\
ametoctradin                                                        & 865318-97-4  \\
amidosulfuron                                                       & 120923-37-7  \\
aminopyralid                                                        & 150114-71-9  \\
amisulbrom                                                          & 348635-87-0  \\
amitrole                                                            & 61-82-5      \\
ammonium acetate                                                    & 631-61-8     \\
ammonium carbonate                                                  & 506-87-6     \\
ammonium thiocyanate                                                & 1762-95-4    \\
ampelomyces quisqualis souche aq10                                  &              \\
anthraquinone                                                       & 84-65-1      \\
asulam sodium                                                       & 2302-17-2    \\
azadirachtin A                                                      & 11141-17-6   \\
azimsulfuron                                                        & 120162-55-2  \\
azoxystrobin                                                        & 131860-33-8  \\
bacillus firmus i-1582                                              &              \\
bacillus pumilus souche qst 2808                                    &              \\
bacillus subtilis                                                   & 68038-70-0   \\
beflubutamid                                                        & 113614-08-7  \\
benalaxyl                                                           & 71626-11-4   \\
benalaxyl-M                                                         & 98243-83-5   \\
benfluralin                                                         & 1861-40-1    \\
benoxacor                                                           & 98730-04-2   \\
bensulfuron-methyl                                                  & 83055-99-6   \\
bentazone                                                           & 25057-89-0   \\
benthiavalicarb isopropyl                                           & 177406-68-7  \\
benzovindiflupyr                                                    & 1072957-71-1 \\
bifenazate                                                          & 149877-41-8  \\
bifenox                                                             & 42576-02-3   \\
bifenthrin                                                          & 82657-04-3   \\
bitertanol                                                          & 55179-31-2   \\
bixafen                                                             & 581809-46-3  \\
boscalid                                                            & 188425-85-6  \\
bromadiolone                                                        & 28772-56-7   \\
bromoxynil                                                          & 1689-84-5    \\
bromoxynil octanoate                                                & 1689-99-2    \\
bromuconazole                                                       & 116255-48-2  \\
bupirimate                                                          & 41483-43-6   \\
calcium cyanamide                                                   & 156-62-7     \\
captan                                                              & 133-06-2     \\
carbendazim                                                         & 10605-21-7   \\
carbetamide                                                         & 16118-49-3   \\
carbofuran                                                          & 1563-66-2    \\
carboxin                                                            & 5234-68-4    \\
carfentrazone-ethyl                                                 & 128639-02-1  \\
cerevisane                                                          &              \\
chlorantraniliprole                                                 & 500008-45-7  \\
chloridazon                                                         & 1698-60-8    \\
chlormequat chloride                                                & 999-81-5     \\
chlorothalonil                                                      & 1897-45-6    \\
chlorotoluron                                                       & 15545-48-9   \\
chlorpropham                                                        & 101-21-3     \\
chlorpyrifos                                                        & 2921-88-2    \\
chlorpyrifos-methyl                                                 & 5598-13-0    \\
choline chloride                                                    & 67-48-1      \\
clethodim                                                           & 99129-21-2   \\
clodinafop-propargyl                                                & 105512-06-9  \\
clomazone                                                           & 81777-89-1   \\
clonostachys rosea j1446 (anciennement   gliocladium catenulatum)   &              \\
clopyralid                                                          & 1702-17-6    \\
clopyralid monoethanolamine salt                                    & 57754-85-5   \\
cloquintocet-mexyl                                                  & 99607-70-2   \\
clothianidin                                                        & 210880-92-5  \\
codlemone                                                           & 33956-49-9   \\
coniothyrium minitans souche con/m/91-08                            &              \\
copper (1) oxide                                                    & 1317-39-1    \\
copper II hydroxide                                                 & 20427-59-2   \\
copper oxychloride                                                  & 1332-40-7    \\
copper sulphate                                                     & 12527-76-3   \\
cos-oga                                                             &              \\
cuivre du tallate de cuivre                                         & 71789-22-8   \\
cyantraniliprole                                                    & 736994-63-1  \\
cyazofamid                                                          & 120116-88-3  \\
cycloxydim                                                          & 101205-02-1  \\
cydia pomonella granulovirus                                        &              \\
cyflufenamid                                                        & 180409-60-3  \\
cyfluthrin                                                          & 68359-37-5   \\
cyhalofop-butyl                                                     & 122008-85-9  \\
cymoxanil                                                           & 57966-95-7   \\
cypermethrin                                                        & 52315-07-8   \\
cyproconazole                                                       & 94361-06-5   \\
cyprodinil                                                          & 121552-61-2  \\
cyprosulfamide                                                      & 221667-31-8  \\
cyromazine                                                          & 66215-27-8   \\
daminozide                                                          & 1596-84-5    \\
dazomet                                                             & 533-74-4     \\
deltamethrin                                                        & 52918-63-5   \\
desmedipham                                                         & 13684-56-5   \\
diazinon                                                            & 333-41-5     \\
dicamba                                                             & 1918-00-9    \\
dichlobenil                                                         & 1194-65-6    \\
dichlorprop-P                                                       & 15165-67-0   \\
diclofop-methyl                                                     & 51338-27-3   \\
dicofol                                                             & 115-32-2     \\
difenoconazole                                                      & 119446-68-3  \\
diflubenzuron                                                       & 35367-38-5   \\
diflufenican                                                        & 83164-33-4   \\
dimethachlor                                                        & 50563-36-5   \\
dimethenamid-P                                                      & 163515-14-8  \\
dimethomorph                                                        & 110488-70-5  \\
dimoxystrobin                                                       & 149961-52-4  \\
diquat dibromide                                                    & 85-00-7      \\
disodium phosphonate                                                & 13708-85-5   \\
dithianon                                                           & 3347-22-6    \\
diuron                                                              & 330-54-1     \\
dodine                                                              & 03-10-2439\\
e8,e10-dodecadiene-1-ol + tetradecylacetate&              \\
emamectin benzoate                                                  & 155569-91-8  \\
esfenvalerate                                                       & 66230-04-4   \\
essence de girofle                                                  &              \\
ethephon                                                            & 16672-87-0   \\
ethofumesate                                                        & 26225-79-6   \\
ethoprophos                                                         & 13194-48-4   \\
etofenprox                                                          & 80844-07-1   \\
etoxazole                                                           & 153233-91-1  \\
eugenol                                                             & 97-53-0      \\
famoxadone                                                          & 131807-57-3  \\
farine de sang                                                      &              \\
fenamidone                                                          & 161326-34-7  \\
fenazaquin                                                          & 120928-09-8  \\
fenbuconazole                                                       & 114369-43-6  \\
fenhexamid                                                          & 126833-17-8  \\
fenitrothion                                                        & 122-14-5     \\
fenoxaprop-P-ethyl                                                  & 71283-80-2   \\
fenpropidin                                                         & 67306-00-7   \\
fenpropimorph                                                       & 67564-91-4   \\
fenpyrazamine                                                       & 473798-59-3  \\
ferric phosphate                                                    & 10045-86-0   \\
flazasulfuron                                                       & 104040-78-0  \\
flonicamid                                                          & 158062-67-0  \\
florasulam                                                          & 145701-23-1  \\
fluazifop-P-butyl                                                   & 79241-46-6   \\
fluazinam                                                           & 79622-59-6   \\
fludioxonil                                                         & 131341-86-1  \\
flufenacet                                                          & 142459-58-3  \\
flumioxazin                                                         & 103361-09-7  \\
fluopicolide                                                        & 239110-15-7  \\
fluopyram                                                           & 658066-35-4  \\
fluoxastrobin                                                       & 361377-29-9  \\
flupyrsulfuron-methyl-sodium                                        & 144740-54-5  \\
flurochloridone                                                     & 61213-25-0   \\
fluroxypyr                                                          & 69377-81-7   \\
flurtamone                                                          & 96525-23-4   \\
flusilazole                                                         & 85509-19-9   \\
flutolanil                                                          & 66332-96-5   \\
flutriafol                                                          & 76674-21-0   \\
fluxapyroxad                                                        & 907204-31-3  \\
folpet                                                              & 133-07-3     \\
foramsulfuron                                                       & 173159-57-4  \\
forchlorfenuron                                                     & 68157-60-8   \\
fosetyl                                                             & 15845-66-6   \\
fosetyl-aluminium                                                   & 39148-24-8   \\
fosthiazate                                                         & 98886-44-3   \\
gamma-cyhalothrin                                                   & 76703-62-3   \\
gibberellic acid                                                    & 77-06-5      \\
glufosinate-ammonium                                                & 77182-82-2   \\
glyphosate                                                          & 1071-83-6    \\
graisse de mouton                                                   &              \\
halauxifen                                                          & 943832-60-8  \\
halosulfuron-methyl                                                 & 100784-20-1  \\
heptamaloxyloglucan                                                 & 870721-81-6  \\
hexythiazox                                                         & 78587-05-0   \\
hydrolysat de proteines                                             &              \\
hymexazol                                                           & 10004-44-1   \\
imazalil                                                            & 35554-44-0   \\
imazamox                                                            & 114311-32-9  \\
imazaquin                                                           & 81335-37-7   \\
imidacloprid                                                        & 138261-41-3  \\
indolylbutyric acid                                                 & 133-32-4     \\
indoxacarb                                                          & 173584-44-6  \\
iodosulfuron-methyl-sodium                                          & 144550-36-7  \\
ioxynil                                                             & 1689-83-4    \\
ioxynil octanoate                                                   & 3861-47-0    \\
ipconazole                                                          & 125225-28-7  \\
iprodione                                                           & 36734-19-7   \\
iprovalicarb                                                        & 140923-17-7  \\
iron sulphate                                                       & 7720-78-7    \\
isofetamid                                                          & 875915-78-9  \\
isoproturon                                                         & 34123-59-6   \\
isoxaben                                                            & 82558-50-7   \\
isoxadifen ethyl                                                    & 163520-33-0  \\
isoxaflutole                                                        & 141112-29-0  \\
kaolin                                                              & 1332-58-7    \\
kieselguhr                                                          &              \\
kresoxim-methyl                                                     & 143390-89-0  \\
lambda-cyhalothrin                                                  & 91465-08-6   \\
laminarin                                                           & 9008-22-4    \\
lenacil                                                             & 2164-08-01   \\
lime sulphur                                                        & 1344-81-6    \\
linuron                                                             & 330-55-2     \\
magnesium phosphide                                                 & 12057-74-8   \\
malathion                                                           & 121-75-5     \\
maleic hydrazide                                                    & 123-33-1     \\
maltodextrin                                                        & 9050-36-6    \\
mancozeb                                                            & 8018-01-7   \\
mandipropamid                                                       & 374726-62-2  \\
maneb                                                               & 12427-38-2   \\
mecoprop                                                            & 7085-19-0    \\
mecoprop-P                                                          & 16484-77-8   \\
mefenpyr diethyl                                                    & 135590-91-9  \\
mepanipyrim                                                         & 110235-47-7  \\
mepiquat chloride                                                   & 24307-26-4   \\
meptyldinocap                                                       & 131-72-6     \\
mesosulfuron-methyl                                                 & 208465-21-8  \\
mesotrione                                                          & 104206-82-8  \\
metalaxyl                                                           & 57837-19-1   \\
metalaxyl-M                                                         & 70630-17-0   \\
metaldehyde                                                         & 108-62-3     \\
metam-sodium                                                        & 137-42-8     \\
metamitron                                                          & 41394-05-2   \\
metazachlor                                                         & 67129-08-2   \\
metconazole                                                         & 125116-23-6  \\
methiocarb                                                          & 2032-65-7    \\
methoxyfenozide                                                     & 161050-58-4  \\
metiram                                                             & 9006-42-2    \\
metobromuron                                                        & 3060-89-7    \\
metrafenone                                                         & 220899-03-6  \\
metribuzin                                                          & 21087-64-9   \\
metsulfuron-methyl                                                  & 74223-64-6   \\
milbemectin                                                         & 51596-10-2   \\
myclobutanil                                                        & 88671-89-0   \\
myristyl alcohol                                                    & 112-72-1     \\
napropamide                                                         & 15299-99-7   \\
nicosulfuron                                                        & 111991-09-4  \\
nucleopolyhedrovirus de spodoptera   littoralis                     &              \\
orange oil                                                          & 8028-48-6    \\
oryzalin                                                            & 19044-88-3   \\
oxadiazon                                                           & 19666-30-9   \\
oxamyl                                                              & 23135-22-0   \\
oxathiapiprolin                                                     & 1003318-67-9 \\
oxyfluorfen                                                         & 42874-03-3   \\
paclobutrazol                                                       & 76738-62-0   \\
paecilomyces fumosoroseus apopka souche   97                        &              \\
paraffin oil (C11-C30) (4c)                                         & 97862-82-3   \\
paraffin oil (C18-C30) (1, not ASU)                                 & 8042-47-5    \\
pelargonic acid                                                     & 112-05-0     \\
penconazole                                                         & 66246-88-6   \\
pencycuron                                                          & 66063-05-6   \\
pendimethalin                                                       & 40487-42-1   \\
penoxsulam                                                          & 219714-96-2  \\
penthiopyrad                                                        & 183675-82-3  \\
pethoxamid                                                          & 106700-29-2  \\
phenmedipham                                                        & 13684-63-4   \\
phlebiopsis gigantea                                                &              \\
phosalone                                                           & 2310-17-0    \\
phosmet                                                             & 732-11-6     \\
picloram                                                            & 1918-02-1   \\
picolinafen                                                         & 137641-05-5  \\
picoxystrobin                                                       & 117428-22-5  \\
pinoxaden                                                           & 243973-20-8  \\
piperonyl butoxide                                                  & 51-03-6      \\
pirimicarb                                                          & 23103-98-2   \\
pirimiphos-methyl                                                   & 29232-93-7   \\
poivre                                                              &              \\
potassium bicarbonate                                               & 298-14-6     \\
prochloraz                                                          & 67747-09-5   \\
prohexadione-calcium                                                & 127277-53-6  \\
propamocarb                                                         & 24579-73-5   \\
propaquizafop                                                       & 111479-05-1  \\
propiconazole                                                       & 60207-90-1   \\
propoxycarbazone-sodium                                             & 181274-15-7  \\
propyzamide                                                         & 23950-58-5   \\
proquinazid                                                         & 189278-12-4  \\
prosulfocarb                                                        & 52888-80-9   \\
prosulfuron                                                         & 94125-34-5   \\
prothioconazole                                                     & 178928-70-6  \\
pseudomonas chlororaphis souche ma 342                              &              \\
pymetrozine                                                         & 123312-89-0  \\
pyraclostrobin                                                      & 175013-18-0  \\
pyraflufen-ethyl                                                    & 129630-19-9  \\
pyrethrins (mix)                                                    & 8003-34-7    \\
pyridate                                                            & 55512-33-9   \\
pyrimethanil                                                        & 53112-28-0   \\
pyriofenone                                                         & 688046-61-9  \\
pyriproxyfen                                                        & 95737-68-1   \\
pyroxsulam                                                          & 422556-08-9  \\
quinmerac                                                           & 90717-03-6   \\
quinoclamine                                                        & 2797-51-5    \\
quinoxyfen                                                          & 124495-18-7  \\
quizalofop-P-ethyl                                                  & 100646-51-3  \\
rimsulfuron                                                         & 122931-48-0  \\
rotenone                                                            & 83-79-4      \\
sedaxane                                                            & 874967-67-6  \\
silthiofam                                                          & 175217-20-6  \\
sintofen                                                            & 130561-48-7  \\
sodium hypochlorite                                                 & 7681-52-9    \\
spearmint oil                                                       & 8008-79-5    \\
spinetoram                                                          & 935545-74-7  \\
spinosad                                                            & 168316-95-8  \\
spirodiclofen                                                       & 148477-71-8  \\
spiromesifen                                                        & 283594-90-1  \\
spirotetramat                                                       & 203313-25-1  \\
spiroxamine                                                         & 118134-30-8  \\
sulcotrione                                                         & 99105-77-8   \\
sulfosulfuron                                                       & 141776-32-1  \\
sulfuryl fluoride                                                   & 2699-79-8    \\
sulphur                                                             & 7704-34-9    \\
tau-fluvalinate                                                     & 102851-06-9  \\
tebuconazole                                                        & 107534-96-3  \\
tebufenozide                                                        & 112410-23-8  \\
tebufenpyrad                                                        & 119168-77-3  \\
tefluthrin                                                          & 79538-32-2   \\
tembotrione                                                         & 335104-84-2  \\
terbuthylazine                                                      & 5915-41-3    \\
tetraconazole                                                       & 112281-77-3  \\
thiabendazole                                                       & 148-79-8     \\
thiacloprid                                                         & 111988-49-9  \\
thiamethoxam                                                        & 153719-23-4  \\
thiencarbazone-methyl                                               & 317815-83-1  \\
thifensulfuron-methyl                                               & 79277-27-3   \\
thiodicarb                                                          & 59669-26-0   \\
thiophanate-methyl                                                  & 23564-05-8   \\
thiram                                                              & 137-26-8     \\
thymol                                                              & 89-83-8      \\
tri-allate                                                          & 2303-17-5    \\
triadimenol                                                         & 55219-65-3   \\
tribenuron-methyl                                                   & 101200-48-0  \\
trichoderma asperellum icc012                                       &              \\
trichoderma asperellum tv1                                          &              \\
trichoderma harzianum souche t22                                    &              \\
triclopyr                                                           & 55335-06-3   \\
trifloxystrobin                                                     & 141517-21-7  \\
trifluralin                                                         & 1582-09-8    \\
triflusulfuron-methyl                                               & 126535-15-7  \\
trimethylamine hydrochloride                                        & 593-81-7     \\
trinexapac-ethyl                                                    & 95266-40-3   \\
triticonazole                                                       & 131983-72-7  \\
tritosulfuron                                                       & 142469-14-5  \\
valifenalate                                                        & 283159-90-0  \\
virus de la granulose                                               &              \\
zinc phosphide                                                      & 1314-84-7    \\
ziram                                                               & 137-30-4     \\
zoxamide                                                            & 156052-68-5 
\label{tab:ASlist}
\end{longtable}
\end{document}